\documentclass[preprint,12pt]{elsarticle}

\newcommand*{\affmark}[1][*]{\textsuperscript{#1}}

\journal{arXiv}

\usepackage{graphicx}
\usepackage{multicol}
\usepackage{mathtools}
\usepackage{amssymb}
\usepackage{tikz}
\usepackage{multirow}

\usepackage{graphicx}        %
\usepackage{multicol}        %
\usepackage[bottom]{footmisc}%
\usepackage{subcaption}
\usepackage{pgfplots}
\usepackage{newtxtext}       %
\usepackage[varvw]{newtxmath}       %
\usepackage{todonotes}
\usepackage[overload]{empheq}
\usepackage{ulem}
\usepackage{amsmath}
\usepackage{pgfplots}
\usepackage{tikz}
\usepackage{pgfplots}
\usepackage{wrapfig}
\usepackage{siunitx}
\usepackage{bm}
\usepackage{mhchem}

\usepackage{xcolor}

\usetikzlibrary{spy,arrows}
\usetikzlibrary{shapes.geometric}
\usetikzlibrary{intersections}
\usepgfplotslibrary{groupplots}
\usepgfplotslibrary{colorbrewer}
\usepgfplotslibrary{fillbetween}
\usetikzlibrary{colorbrewer}
\usetikzlibrary{external}
\usetikzlibrary{spy,calc}
\usetikzlibrary{decorations.pathreplacing}
\usetikzlibrary{3d}
\usetikzlibrary{spy,arrows}
\usetikzlibrary{shapes.geometric}
\usetikzlibrary{intersections}
\usetikzlibrary{calc}
\tikzstyle{rounded}=[rounded corners]     
\tikzstyle{function} = [rectangle, text centered, draw=black,minimum width=6.0cm,minimum height=0.7cm]
\tikzstyle{decision} = [rectangle, rounded corners, text centered, draw=black,minimum width=4.0cm,minimum height=0.7cm]
\tikzstyle{module} = [rectangle, rounded corners, text centered, draw=black, double, minimum width=4.0cm,minimum height=0.7cm]
\usepackage{mathrsfs}

\definecolor{grey}{RGB}{230.,230.,230}
\definecolor{lightgrey}{RGB}{240.,240.,250}
\definecolor{lightblue}{RGB}{80, 200,255}
\definecolor{whiteblue}{RGB}{120, 220,255}
\definecolor{darkblue}{RGB}{20, 200, 200}
\definecolor{lighturkis}{RGB}{42, 220,255}
\definecolor{whitegrey}{RGB}{200, 200,220}

\newcommand{\uL}{\mathbf{u}_1}
\newcommand{\uR}{\mathbf{u}_2}
\newcommand{\uBZ}{\mathbf{u}_0}
\newcommand{\UB}{\boldsymbol{U}}

\newcommand{\DGFV}{DG/FV }
\newcommand{\DG}{DG }
\newcommand{\FV}{FV }

\newcommand{\uB}{\boldsymbol{u}}
\newcommand{\FB}{\boldsymbol{F}}
\newcommand{\vB}{\boldsymbol{v}}
\newcommand{\xB}{\boldsymbol{x}}
\newcommand{\nB}{\boldsymbol{n}}
\newcommand{\VB}{\boldsymbol{V}}

\newcommand{\MB}{\boldsymbol{M}}
\newcommand{\xiB}{\boldsymbol{\xi}}
\newcommand{\FtB}{\boldsymbol{\mathcal{F}}}

\newcommand{\DGFVB}{\boldsymbol{V}^{\mathrm{DG}\rightsquigarrow\mathrm{FV}}}
\newcommand{\FVDGB}{\boldsymbol{V}^{\mathrm{FV}\rightsquigarrow\mathrm{DG}}}
\newcommand{\NTOM}{\boldsymbol{V}^{\mathrm{N}\rightsquigarrow\mathrm{\tilde{N}}}}
\newcommand{\MTON}{\boldsymbol{V}^{\mathrm{\tilde{N}}\rightsquigarrow\mathrm{N}}}

\newcommand{\QB}{\boldsymbol{Q}}

\newcommand{\qB}{\boldsymbol{q}}
\newcommand{\gB}{\boldsymbol{g}}

\newcommand{\JB}{\boldsymbol{J}}

\newcommand{\DB}{\boldsymbol{D}}
\newcommand{\YB}{\boldsymbol{Y}}

\newcommand{\xGP}{-0.866, -0.339,  0.339, 0.866}

\newcommand{\xGPI}{-0.774597, 0.,  0.774597}
\newcommand{\xFV}{-0.8, -0.4,0,0.4,0.8}
\newcommand{\innerborders}{-0.6,-0.2,0.2,0.6}

\definecolor{dgcolor}{RGB}{0,200,240}
\definecolor{fvcolor}{RGB}{230,0,150}

\pgfplotsset{
	compat=1.16,
	cycle list/Dark2,
	siunitxlabels/.style={
		/pgfplots/typeset ticklabel/.code={{\pgfmathparse{\tick}$\num[zero-decimal-to-integer]{\pgfmathresult}$}},
	},
	  colormap/hp_convergence/.style={colormap={hp_convergence}{
		rgb=(1.0,0.0,0.53);
		rgb=(0.0,0.9294,1.0);
	  }
	},
	  colormap/hp_freestream/.style={colormap={hp_freestream}{
		rgb=(0.219607843137255,	0.219607843137255	,0.219607843137255);
		rgb=(0.815686274509804,	1	                ,0.952941176470588);
		rgb=(0.380392156862745,	0.854901960784314	,1);
		rgb=(0.364705882352941,	0.250980392156863	,1);
		rgb=(0.811764705882353,	0.133333333333333	,1);
	  }
	},
	colormap/hp_shock_triple/.style={colormap={hp_shock_triple}{
		rgb=(0.219607843137255,	0.219607843137255,	0.219607843137255);
		rgb=(0.819607843137255,	0.862745098039216,	0.850980392156863);
		rgb=(0.407843137254902,	0.568627450980392,	0.850980392156863);
		rgb=(0.352941176470588,	0.686274509803922,	0.713725490196078);
	  }
	},
	  colormap/blue_bright/.style={colormap={bright_blue}{
		rgb=(0.015686274509804,	0.015686274509804,	0.015686274509804);
		rgb=(0.149019607843137,	0.145098039215686,	0.168627450980392);
		rgb=(0.23921568627451,	0.231372549019608,	0.305882352941176);
		rgb=(0.290196078431373,	0.282352941176471,	0.431372549019608);
		rgb=(0.294117647058823,	0.376470588235294,	0.588235294117647);
		rgb=(0.329411764705882,	0.607843137254902,	0.768627450980392);
		rgb=(0.55686274509803,0.8,0.854901960784314);
		rgb=(1,1,1);
	  }
	},
	colormap/bright_blue/.style={colormap={blue_bright}{
		rgb=(1,1,1);
		rgb=(0.55686274509803,0.8,0.854901960784314);
		rgb=(0.329411764705882,	0.607843137254902,	0.768627450980392);
		rgb=(0.294117647058823,	0.376470588235294,	0.588235294117647);
		rgb=(0.290196078431373,	0.282352941176471,	0.431372549019608);
		rgb=(0.23921568627451,	0.231372549019608,	0.305882352941176);
		rgb=(0.149019607843137,	0.145098039215686,	0.168627450980392);
		rgb=(0.015686274509804,	0.015686274509804,	0.015686274509804);
	  }
	},
	colormap/jet_species_map/.style={colormap={jet_species_map}{
		rgb=(0.0, 0.0, 0.0);
		rgb=(0.8, 0.8, 0.8);
		rgb=(0.5, 0.5, 0.9);
		rgb=(0.3, 0.0, 0.4);
	  }
	},
	colormap/jet_pressure_map/.style={colormap={jet_pressure_map}{
		rgb=(1.0, 1.0, 1.0);
		rgb=(0.4, 0.5, 0.7);
		rgb=(0.0, 0.0, 0.0);
	  }
	},
		colormap/coolwarm_ext/.style={colormap={coolwarm_ext}{
		rgb=(0.0, 0.0, 0.34902);
		rgb=(0.039216, 0.062745, 0.380392);
		rgb=(0.062745, 0.117647, 0.411765);
		rgb=(0.090196, 0.184314, 0.45098);
		rgb=(0.12549, 0.262745, 0.501961);
		rgb=(0.160784, 0.337255, 0.541176);
		rgb=(0.2, 0.396078, 0.568627);
		rgb=(0.239216, 0.454902, 0.6);
		rgb=(0.286275, 0.521569, 0.65098);
		rgb=(0.337255, 0.592157, 0.701961);
		rgb=(0.388235, 0.654902, 0.74902);
		rgb=(0.466667, 0.737255, 0.819608);
		rgb=(0.572549, 0.819608, 0.878431);
		rgb=(0.654902, 0.866667, 0.909804);
		rgb=(0.752941, 0.917647, 0.941176);
		rgb=(0.823529, 0.956863, 0.968627);
		rgb=(0.988235, 0.960784, 0.901961);
		rgb=(0.941176, 0.984314, 0.988235);
		rgb=(0.988235, 0.945098, 0.85098);
		rgb=(0.980392, 0.898039, 0.784314);
		rgb=(0.968627, 0.835294, 0.698039);
		rgb=(0.94902, 0.733333, 0.588235);
		rgb=(0.929412, 0.65098, 0.509804);
		rgb=(0.909804, 0.564706, 0.435294);
		rgb=(0.878431, 0.458824, 0.352941);
		rgb=(0.839216, 0.388235, 0.286275);
		rgb=(0.760784, 0.294118, 0.211765);
		rgb=(0.701961, 0.211765, 0.168627);
		rgb=(0.65098, 0.156863, 0.129412);
		rgb=(0.6, 0.094118, 0.094118);
		rgb=(0.54902, 0.066667, 0.098039);
		rgb=(0.501961, 0.05098, 0.12549);
		rgb=(0.45098, 0.054902, 0.172549);
		rgb=(0.4, 0.054902, 0.192157);
		rgb=(0.34902, 0.070588, 0.211765);
		}
	},
	colormap/hp_comparison/.style={colormap={hp_comparison}{
		rgb=(0.854901960784314,	0.898039215686275,	0.91372549019607);
		rgb=(0.686274509803922,	1,	0.980392156862745);
		rgb=(0.36078431372549,	0.580392156862745,	0.862745098039216);
		rgb=(0.345098039215686,	0.368627450980392,	0.411764705882353);
	}
	},
	colormap/XRay/.style={colormap={XRay}{
			rgb=(1.0, 1.0, 1.0);
			rgb=(0.0, 0.0, 0.0);
		}
	},
	colormap/inferno/.style={colormap={inferno}{
		rgb=(0.001462, 0.000466, 0.013866);
		rgb=(0.002267, 0.001270, 0.018570);
		rgb=(0.003299, 0.002249, 0.024239);
		rgb=(0.004547, 0.003392, 0.030909);
		rgb=(0.006006, 0.004692, 0.038558);
		rgb=(0.007676, 0.006136, 0.046836);
		rgb=(0.009561, 0.007713, 0.055143);
		rgb=(0.011663, 0.009417, 0.063460);
		rgb=(0.013995, 0.011225, 0.071862);
		rgb=(0.016561, 0.013136, 0.080282);
		rgb=(0.019373, 0.015133, 0.088767);
		rgb=(0.022447, 0.017199, 0.097327);
		rgb=(0.025793, 0.019331, 0.105930);
		rgb=(0.029432, 0.021503, 0.114621);
		rgb=(0.033385, 0.023702, 0.123397);
		rgb=(0.037668, 0.025921, 0.132232);
		rgb=(0.042253, 0.028139, 0.141141);
		rgb=(0.046915, 0.030324, 0.150164);
		rgb=(0.051644, 0.032474, 0.159254);
		rgb=(0.056449, 0.034569, 0.168414);
		rgb=(0.061340, 0.036590, 0.177642);
		rgb=(0.066331, 0.038504, 0.186962);
		rgb=(0.071429, 0.040294, 0.196354);
		rgb=(0.076637, 0.041905, 0.205799);
		rgb=(0.081962, 0.043328, 0.215289);
		rgb=(0.087411, 0.044556, 0.224813);
		rgb=(0.092990, 0.045583, 0.234358);
		rgb=(0.098702, 0.046402, 0.243904);
		rgb=(0.104551, 0.047008, 0.253430);
		rgb=(0.110536, 0.047399, 0.262912);
		rgb=(0.116656, 0.047574, 0.272321);
		rgb=(0.122908, 0.047536, 0.281624);
		rgb=(0.129285, 0.047293, 0.290788);
		rgb=(0.135778, 0.046856, 0.299776);
		rgb=(0.142378, 0.046242, 0.308553);
		rgb=(0.149073, 0.045468, 0.317085);
		rgb=(0.155850, 0.044559, 0.325338);
		rgb=(0.162689, 0.043554, 0.333277);
		rgb=(0.169575, 0.042489, 0.340874);
		rgb=(0.176493, 0.041402, 0.348111);
		rgb=(0.183429, 0.040329, 0.354971);
		rgb=(0.190367, 0.039309, 0.361447);
		rgb=(0.197297, 0.038400, 0.367535);
		rgb=(0.204209, 0.037632, 0.373238);
		rgb=(0.211095, 0.037030, 0.378563);
		rgb=(0.217949, 0.036615, 0.383522);
		rgb=(0.224763, 0.036405, 0.388129);
		rgb=(0.231538, 0.036405, 0.392400);
		rgb=(0.238273, 0.036621, 0.396353);
		rgb=(0.244967, 0.037055, 0.400007);
		rgb=(0.251620, 0.037705, 0.403378);
		rgb=(0.258234, 0.038571, 0.406485);
		rgb=(0.264810, 0.039647, 0.409345);
		rgb=(0.271347, 0.040922, 0.411976);
		rgb=(0.277850, 0.042353, 0.414392);
		rgb=(0.284321, 0.043933, 0.416608);
		rgb=(0.290763, 0.045644, 0.418637);
		rgb=(0.297178, 0.047470, 0.420491);
		rgb=(0.303568, 0.049396, 0.422182);
		rgb=(0.309935, 0.051407, 0.423721);
		rgb=(0.316282, 0.053490, 0.425116);
		rgb=(0.322610, 0.055634, 0.426377);
		rgb=(0.328921, 0.057827, 0.427511);
		rgb=(0.335217, 0.060060, 0.428524);
		rgb=(0.341500, 0.062325, 0.429425);
		rgb=(0.347771, 0.064616, 0.430217);
		rgb=(0.354032, 0.066925, 0.430906);
		rgb=(0.360284, 0.069247, 0.431497);
		rgb=(0.366529, 0.071579, 0.431994);
		rgb=(0.372768, 0.073915, 0.432400);
		rgb=(0.379001, 0.076253, 0.432719);
		rgb=(0.385228, 0.078591, 0.432955);
		rgb=(0.391453, 0.080927, 0.433109);
		rgb=(0.397674, 0.083257, 0.433183);
		rgb=(0.403894, 0.085580, 0.433179);
		rgb=(0.410113, 0.087896, 0.433098);
		rgb=(0.416331, 0.090203, 0.432943);
		rgb=(0.422549, 0.092501, 0.432714);
		rgb=(0.428768, 0.094790, 0.432412);
		rgb=(0.434987, 0.097069, 0.432039);
		rgb=(0.441207, 0.099338, 0.431594);
		rgb=(0.447428, 0.101597, 0.431080);
		rgb=(0.453651, 0.103848, 0.430498);
		rgb=(0.459875, 0.106089, 0.429846);
		rgb=(0.466100, 0.108322, 0.429125);
		rgb=(0.472328, 0.110547, 0.428334);
		rgb=(0.478558, 0.112764, 0.427475);
		rgb=(0.484789, 0.114974, 0.426548);
		rgb=(0.491022, 0.117179, 0.425552);
		rgb=(0.497257, 0.119379, 0.424488);
		rgb=(0.503493, 0.121575, 0.423356);
		rgb=(0.509730, 0.123769, 0.422156);
		rgb=(0.515967, 0.125960, 0.420887);
		rgb=(0.522206, 0.128150, 0.419549);
		rgb=(0.528444, 0.130341, 0.418142);
		rgb=(0.534683, 0.132534, 0.416667);
		rgb=(0.540920, 0.134729, 0.415123);
		rgb=(0.547157, 0.136929, 0.413511);
		rgb=(0.553392, 0.139134, 0.411829);
		rgb=(0.559624, 0.141346, 0.410078);
		rgb=(0.565854, 0.143567, 0.408258);
		rgb=(0.572081, 0.145797, 0.406369);
		rgb=(0.578304, 0.148039, 0.404411);
		rgb=(0.584521, 0.150294, 0.402385);
		rgb=(0.590734, 0.152563, 0.400290);
		rgb=(0.596940, 0.154848, 0.398125);
		rgb=(0.603139, 0.157151, 0.395891);
		rgb=(0.609330, 0.159474, 0.393589);
		rgb=(0.615513, 0.161817, 0.391219);
		rgb=(0.621685, 0.164184, 0.388781);
		rgb=(0.627847, 0.166575, 0.386276);
		rgb=(0.633998, 0.168992, 0.383704);
		rgb=(0.640135, 0.171438, 0.381065);
		rgb=(0.646260, 0.173914, 0.378359);
		rgb=(0.652369, 0.176421, 0.375586);
		rgb=(0.658463, 0.178962, 0.372748);
		rgb=(0.664540, 0.181539, 0.369846);
		rgb=(0.670599, 0.184153, 0.366879);
		rgb=(0.676638, 0.186807, 0.363849);
		rgb=(0.682656, 0.189501, 0.360757);
		rgb=(0.688653, 0.192239, 0.357603);
		rgb=(0.694627, 0.195021, 0.354388);
		rgb=(0.700576, 0.197851, 0.351113);
		rgb=(0.706500, 0.200728, 0.347777);
		rgb=(0.712396, 0.203656, 0.344383);
		rgb=(0.718264, 0.206636, 0.340931);
		rgb=(0.724103, 0.209670, 0.337424);
		rgb=(0.729909, 0.212759, 0.333861);
		rgb=(0.735683, 0.215906, 0.330245);
		rgb=(0.741423, 0.219112, 0.326576);
		rgb=(0.747127, 0.222378, 0.322856);
		rgb=(0.752794, 0.225706, 0.319085);
		rgb=(0.758422, 0.229097, 0.315266);
		rgb=(0.764010, 0.232554, 0.311399);
		rgb=(0.769556, 0.236077, 0.307485);
		rgb=(0.775059, 0.239667, 0.303526);
		rgb=(0.780517, 0.243327, 0.299523);
		rgb=(0.785929, 0.247056, 0.295477);
		rgb=(0.791293, 0.250856, 0.291390);
		rgb=(0.796607, 0.254728, 0.287264);
		rgb=(0.801871, 0.258674, 0.283099);
		rgb=(0.807082, 0.262692, 0.278898);
		rgb=(0.812239, 0.266786, 0.274661);
		rgb=(0.817341, 0.270954, 0.270390);
		rgb=(0.822386, 0.275197, 0.266085);
		rgb=(0.827372, 0.279517, 0.261750);
		rgb=(0.832299, 0.283913, 0.257383);
		rgb=(0.837165, 0.288385, 0.252988);
		rgb=(0.841969, 0.292933, 0.248564);
		rgb=(0.846709, 0.297559, 0.244113);
		rgb=(0.851384, 0.302260, 0.239636);
		rgb=(0.855992, 0.307038, 0.235133);
		rgb=(0.860533, 0.311892, 0.230606);
		rgb=(0.865006, 0.316822, 0.226055);
		rgb=(0.869409, 0.321827, 0.221482);
		rgb=(0.873741, 0.326906, 0.216886);
		rgb=(0.878001, 0.332060, 0.212268);
		rgb=(0.882188, 0.337287, 0.207628);
		rgb=(0.886302, 0.342586, 0.202968);
		rgb=(0.890341, 0.347957, 0.198286);
		rgb=(0.894305, 0.353399, 0.193584);
		rgb=(0.898192, 0.358911, 0.188860);
		rgb=(0.902003, 0.364492, 0.184116);
		rgb=(0.905735, 0.370140, 0.179350);
		rgb=(0.909390, 0.375856, 0.174563);
		rgb=(0.912966, 0.381636, 0.169755);
		rgb=(0.916462, 0.387481, 0.164924);
		rgb=(0.919879, 0.393389, 0.160070);
		rgb=(0.923215, 0.399359, 0.155193);
		rgb=(0.926470, 0.405389, 0.150292);
		rgb=(0.929644, 0.411479, 0.145367);
		rgb=(0.932737, 0.417627, 0.140417);
		rgb=(0.935747, 0.423831, 0.135440);
		rgb=(0.938675, 0.430091, 0.130438);
		rgb=(0.941521, 0.436405, 0.125409);
		rgb=(0.944285, 0.442772, 0.120354);
		rgb=(0.946965, 0.449191, 0.115272);
		rgb=(0.949562, 0.455660, 0.110164);
		rgb=(0.952075, 0.462178, 0.105031);
		rgb=(0.954506, 0.468744, 0.099874);
		rgb=(0.956852, 0.475356, 0.094695);
		rgb=(0.959114, 0.482014, 0.089499);
		rgb=(0.961293, 0.488716, 0.084289);
		rgb=(0.963387, 0.495462, 0.079073);
		rgb=(0.965397, 0.502249, 0.073859);
		rgb=(0.967322, 0.509078, 0.068659);
		rgb=(0.969163, 0.515946, 0.063488);
		rgb=(0.970919, 0.522853, 0.058367);
		rgb=(0.972590, 0.529798, 0.053324);
		rgb=(0.974176, 0.536780, 0.048392);
		rgb=(0.975677, 0.543798, 0.043618);
		rgb=(0.977092, 0.550850, 0.039050);
		rgb=(0.978422, 0.557937, 0.034931);
		rgb=(0.979666, 0.565057, 0.031409);
		rgb=(0.980824, 0.572209, 0.028508);
		rgb=(0.981895, 0.579392, 0.026250);
		rgb=(0.982881, 0.586606, 0.024661);
		rgb=(0.983779, 0.593849, 0.023770);
		rgb=(0.984591, 0.601122, 0.023606);
		rgb=(0.985315, 0.608422, 0.024202);
		rgb=(0.985952, 0.615750, 0.025592);
		rgb=(0.986502, 0.623105, 0.027814);
		rgb=(0.986964, 0.630485, 0.030908);
		rgb=(0.987337, 0.637890, 0.034916);
		rgb=(0.987622, 0.645320, 0.039886);
		rgb=(0.987819, 0.652773, 0.045581);
		rgb=(0.987926, 0.660250, 0.051750);
		rgb=(0.987945, 0.667748, 0.058329);
		rgb=(0.987874, 0.675267, 0.065257);
		rgb=(0.987714, 0.682807, 0.072489);
		rgb=(0.987464, 0.690366, 0.079990);
		rgb=(0.987124, 0.697944, 0.087731);
		rgb=(0.986694, 0.705540, 0.095694);
		rgb=(0.986175, 0.713153, 0.103863);
		rgb=(0.985566, 0.720782, 0.112229);
		rgb=(0.984865, 0.728427, 0.120785);
		rgb=(0.984075, 0.736087, 0.129527);
		rgb=(0.983196, 0.743758, 0.138453);
		rgb=(0.982228, 0.751442, 0.147565);
		rgb=(0.981173, 0.759135, 0.156863);
		rgb=(0.980032, 0.766837, 0.166353);
		rgb=(0.978806, 0.774545, 0.176037);
		rgb=(0.977497, 0.782258, 0.185923);
		rgb=(0.976108, 0.789974, 0.196018);
		rgb=(0.974638, 0.797692, 0.206332);
		rgb=(0.973088, 0.805409, 0.216877);
		rgb=(0.971468, 0.813122, 0.227658);
		rgb=(0.969783, 0.820825, 0.238686);
		rgb=(0.968041, 0.828515, 0.249972);
		rgb=(0.966243, 0.836191, 0.261534);
		rgb=(0.964394, 0.843848, 0.273391);
		rgb=(0.962517, 0.851476, 0.285546);
		rgb=(0.960626, 0.859069, 0.298010);
		rgb=(0.958720, 0.866624, 0.310820);
		rgb=(0.956834, 0.874129, 0.323974);
		rgb=(0.954997, 0.881569, 0.337475);
		rgb=(0.953215, 0.888942, 0.351369);
		rgb=(0.951546, 0.896226, 0.365627);
		rgb=(0.950018, 0.903409, 0.380271);
		rgb=(0.948683, 0.910473, 0.395289);
		rgb=(0.947594, 0.917399, 0.410665);
		rgb=(0.946809, 0.924168, 0.426373);
		rgb=(0.946392, 0.930761, 0.442367);
		rgb=(0.946403, 0.937159, 0.458592);
		rgb=(0.946903, 0.943348, 0.474970);
		rgb=(0.947937, 0.949318, 0.491426);
		rgb=(0.949545, 0.955063, 0.507860);
		rgb=(0.951740, 0.960587, 0.524203);
		rgb=(0.954529, 0.965896, 0.540361);
		rgb=(0.957896, 0.971003, 0.556275);
		rgb=(0.961812, 0.975924, 0.571925);
		rgb=(0.966249, 0.980678, 0.587206);
		rgb=(0.971162, 0.985282, 0.602154);
		rgb=(0.976511, 0.989753, 0.616760);
		rgb=(0.982257, 0.994109, 0.631017);
		rgb=(0.988362, 0.998364, 0.644924);
		}
	  },
}

\begin{document}

\begin{frontmatter}

\title{Tackling Compressible Turbulent Multi-Component Flows with Dynamic hp-Adaptation}

\author[]{Pascal Mossier \protect\affmark[a,$*$]}
\ead{pascal.mossier@iag.uni-stuttgart.de}
\author[]{Philipp Oestringer \affmark[a]}
\author[]{Jens Keim \affmark[a]}
\author[]{Catherine Mavriplis\affmark[b]}
\author[]{Andrea D. Beck \affmark[a]}
\author[]{Claus-Dieter Munz \affmark[a]}

\affiliation[1]{organization={Institute of Aerodynamics and Gas Dynamics, University of Stuttgart},
            addressline={Pfaffenwaldring 21}, 
            city={Stuttgart},
            postcode={70569}, 
            country={Germany}}

\affiliation[2]{organization={Department of Mechanical Engineering, University of Ottawa},
			city={Ottawa},
			postcode={K1N 6N5}, 
			country={Canada}}
\begin{abstract}
	In this paper, we present an hp-adaptive hybrid Discontinuous Galerkin/Finite Volume method for simulating 
	compressible, turbulent multi-component flows. Building on a previously established hp-adaptive strategy for hyperbolic
	gas- and droplet-dynamics problems, this study extends the hybrid \DGFV approach to viscous flows with multiple species 
	and incorporates non-conforming interfaces, enabling enhanced flexibility in grid generation. A central contribution of this work 
	lies in the computation of both convective and dissipative fluxes across non-conforming element interfaces 
	of mixed discretizations. To achieve accurate shock localization and scale-resolving representation of turbulent structures, 
	the operator dynamically switches between an h-refined \FV sub-cell scheme and a p-adaptive \DG method, based on an a priori 
	modal solution analysis. The method is implemented in the high-order open-source framework FLEXI and validated against 
	benchmark problems, including the supersonic Taylor-Green vortex and a triplepoint shock interaction, demonstrating 
	its robustness and accuracy for under-resolved shock-turbulence interactions and compressible multi-species scenarios. 
	Finally, the method's capabilities are showcased through 
	an implicit large eddy simulation of an under-expanded hydrogen jet mixing with air, highlighting its potential for 
	tackling challenging compressible multi-species flows in engineering.
\end{abstract}			

\begin{keyword}
hp-Refinement \sep Multi-Species Navier--Stokes \sep Non-Conforming Meshes \sep Supersonic Jet Flow \sep Compressible Large Eddy-Simulation 
\end{keyword}

\end{frontmatter}

\newpage
\section{Introduction}
\label{sec:Introduction}
Turbulent multi-component flows are relevant to a wide range of engineering applications. A prominent example 
in the face of the climate crisis and the depletion of fossil fuel reserves is the adaptation of internal
combustion engines to hydrogen. Here, a close understanding and accurate prediction of the high-speed
injection and mixing of the hydrogen with air plays a critical role in controlling the combustion process.
Simulating such flows presents significant challenges due to the nonlinear interactions of shock waves, turbulence, 
acoustics, and material interfaces, which span a wide range of spatial and temporal scales. 

High-order methods are widely recognized as adept and efficient tools for scale-resolving 
computations of turbulent flows, due to their exponential error convergence for smooth problems. 
Prominent examples are Finite Volume (FV) methods with weighted essentially
non-oscillatory (WENO)~\cite{Liu1994,Shu1998} and central weighted essentially non-oscillatory (CWENO)~\cite{Dumbser2017} reconstructions, 
spectral difference methods~\cite{Liu2006} or flux reconstruction methods~\cite{Huynh2007}. 
A particularly efficient approach is the Discontinuous Galerkin Spectral Element Method (DGSEM)~\cite{Kopriva2009}, 
which leverages its tensor basis structure and high data locality, to allow for comparatively easy implementation and 
exceptional scalability on massively parallel architectures. 

However, compressible multi-species flows encounter a long-standing weakness of \DG methods, namely their tendency 
to produce spurious Gibbs-oscillations in the presence of non-linear fluxes and discontinuous solutions. 
There exists a panoply of strategies to cope with this issue in the literature, ranging from the addition
of artificial viscosity~\cite{Neumann1950,Persson2013,Zeifang2021}, to filtering~\cite{Krivodonova2007,Shu2010}, flux reconstruction~\cite{Vilar2019} and the inclusion 
of piece-wise constant ansatz functions~\cite{Huerta2012}. 

The latter approach allows for recovery of sub-element information
through the inclusion of a Finite Volume sub-cell grid, as proposed by~\cite{Persson2013}. During recent years, different flavors of this 
hybrid \DGFV discretization have emerged: Sonntag and Munz presented an a priori switching between \DG and \FV sub-cell
discretization with a common number of degrees of freedom (DOFs) per element. Dumbser and Loub\`{e}re developed the 
multi-dimensional optimal order detection (MOOD) strategy~\cite{Dumbser2016a}, which replaces a \DG candidate solution in
an a posteriori fashion with a WENO sub-cell reconstruction of decreasing order until an admissible solution is found. 
Finally, Hennemann et al.~\cite{Hennemann2021} proposed a convex combination 
of element local \DG and \FV solutions, allowing for a gradual transition between both operators.  

This paper builds on the work of Mossier et al.~\cite{Mossier2022,Mossier2023,MossierPHD}, who introduced an 
hp-adaptive extension to the \FV sub-cell implementation of Sonntag and Munz~\cite{Sonntag2014,Sonntag2017}.
The method enables a finer \FV sub-cell resolution, decoupled from the \DG ansatz degree, effectively mitigating
accuracy degradation from order reduction through sub-cell h-refinement. 
In combination with a dynamic adaptation of the \DG ansatz degree, an hp-adaptive hybrid \DGFV operator
was constructed. It takes advantage of the exponential convergence of the p-adaptive \DG operator in smooth regions, as well as 
the excellent shock localization of the robust \FV discretization on an h-refined sub-cell grid.
The resulting hp-adaptive scheme was validated for
gas dynamics benchmarks in~\cite{Mossier2022} and implemented in a sharp-interface framework, where it 
proved to be well suited for challenging droplet-dynamics problems~\cite{Mossier2022,Mossier2023,Mossier2025}.

With the present work, the hybrid hp-adaptive \DGFV operator is extended to viscous problems with multiple species 
to study compressible, turbulent multi-component flows. Since mesh generation with hexahedral elements remains challenging,
the paper generalizes element couplings to non-conforming interfaces, facilitating flexible grid generation with local refinement. 
A key challenge in this context is the computation of convective and dissipative fluxes at non-conforming element interfaces of mixed discretizations.
The scheme is implemented as an extension to the open-source high-order code framework FLEXI~\cite{flexi}. 

We assess the performance of the novel method in handling under-resolved turbulence in the presence of
shocks with the supersonic Taylor-Green vortex benchmark~\cite{Chapelier2024}, and demonstrate its multi-species capabilities on
non-conforming grids through a triplepoint shock interaction problem. 
The scheme is finally applied to an implicit large eddy simulation (LES) of an under-expanded $\text{H}_2$-jet, immersed in an 
air atmosphere and compared to studies of Hamzehloo~\cite{Hamzehloo2014} and Vuorinen~\cite{Vuorinen2013}.

The paper is organized as follows: Section \ref{sec:Equations} revisits the governing continuum equations for compressible multi-species flows.
In Section \ref{sec:Numerics}, the p-adaptive DGSEM discretization and \FV sub-cell method are derived for hyperbolic-parabolic conservation
equations with both convective and dissipative fluxes. 
The section focuses on the flux computation at non-conforming interfaces of variable discretizations and outlines the temporal 
discretization and the indicator scheme for multi-component flows. 
Further, non-linear stability is addressed with a split-form extension of the DGSEM and a positivity preserving limiter.
The scheme is validated with free-stream and convergence studies in Section \ref{sec:Validation} and applied to a compressible
Taylor-Green-Vortex and a shock-triplepoint interaction. Finally, Section \ref{sec:Application} presents a jet simulation 
involving $\text{H}_2$ and air mixing under compressible conditions. The paper concludes with a summary and discussion in Section \ref{sec:Conclusion}.

\section{Governing Equations}
\label{sec:Equations}
In the present study, the Navier--Stokes equations are applied to model compressible, turbulent, multi-component flows.
We consider a computational domain $\Omega$ bounded by $\Gamma=\partial \Omega$, a physical coordinate vector 
$\xB=\left(x_1,x_2,x_3\right)^T\in\Omega$ and a time interval $(0,t]$.
Assuming a fluid with $N_k$ species, the Navier--Stokes equations are defined as
\begin{subequations}
	\begin{align}
		\frac{\partial \rho}{\partial t} & + \nabla_{\xB} \cdot (\rho \vB) = 0, \\
		\frac{\partial \rho \vB}{\partial t} & + \nabla_{\xB} \cdot \left(\rho \vB \otimes \vB + p \mathbb{I}\right)= \nabla_{\xB} \cdot \left(\boldsymbol{\tau}\right), \\
		\frac{\partial \rho e} {\partial t} & + \nabla_{\xB} \cdot \left[\left(\rho e + p \mathbb{I}\right)\vB\right]= \nabla_{\xB} \cdot \left(\boldsymbol{\tau}\cdot\vB-\qB_h-\qB_d\right), \\
		\frac{\partial \rho Y_k}{\partial t} & + \nabla_{\xB} \cdot (\rho Y_k \vB) = \nabla_{\xB}\cdot\left(-\JB_k\right), \quad k=1,...,N_k-1,
	\end{align}
	\label{eq:ns_equation}
\end{subequations}
and can be expressed in flux notation as 
\begin{subequations}
\begin{align}
	\frac{\partial \uB}{\partial t}+\nabla_{\xB} \cdot \FB^c\left(\uB\right)+\nabla_{\xB} \cdot \FB^v\left(\uB\right)&=0 \\
	\Leftrightarrow \frac{\partial \uB}{\partial t}+\nabla_{\xB} \cdot \left(\FB\left(\uB,\nabla_{\xB} \uB\right) \right)&=0
	\label{eq:ns_flux_notation}
\end{align}
\end{subequations}
with the conservative state vector $\uB=\left(\rho,\rho\vB,\rho e, \rho \YB \right)$, the convective flux $\FB_v$ and the viscous flux $\FB_v$ in terms of the 
density $\rho$, the velocity vector $\vB=\left(v_1,v_2,v_3\right)^T$, the mass-specific total energy $e$, the pressure $p$
and the vector of mass fractions $\YB=\left(Y_1,...,Y_{N_k-1}\right)^T$. 
It is sufficient to consider the first $N_k-1$ evolution equations for the mass fractions $Y_k=\frac{\rho_k}{\rho}$ since the relationships
\begin{equation}
	\sum_{j=1}^{N_k} Y_j = 1,   \quad \sum_{j=1}^{N_k} \rho_j = \rho
	\label{eq:species_closure}
\end{equation}
have to hold for the sum of the mass fractions and partial densities. 
Assuming a Newtonian fluid of multiple species $N_k>1$, Fourier's hypothesis for heat conduction and Fickian diffusion, 
the stress tensor $\boldsymbol{\tau}$, heat flux $\qB_h$ and interspecies enthalpy flux $\qB_d$
are given as 
\begin{subequations}
	\begin{align}
		\boldsymbol{\tau}&=\mu\left[\nabla_{\xB}\vB+\nabla_{\xB}\vB^T-\frac{2}{3}\left(\nabla_{\xB}\cdot \vB\right)\mathbb{I}\right], \label{eq:visc_stress}\\
		\qB_h&=-\lambda \nabla_{\xB} T, \label{eq:fourier_law}\\
		\qB_d&=\sum_{j}h_j \JB_j, \label{eq:enthalpy_flux}
	\end{align}
\end{subequations}
with the dynamic viscosity $\mu$, the heat conductivity $\lambda$ and the species diffusion flux $\JB_k$ given as
\begin{equation}
\JB_k=-\rho D_k\nabla_{\xB} Y_k - \rho Y_k\sum_{j}^{N_{k}-1}D_j\nabla_{\xB} Y_j , \label{eq:fickian_law}
\end{equation}
where $D_k$ denotes the diffusion coefficient. The second term on the right side of 
Equation \eqref{eq:fickian_law} is a corrective term to recover $\sum_{j}\JB_j=0$ and thus ensure local mass conservation~\cite{Coffee1981}. 

The mass-specific total energy $e$ comprises a specific internal energy $\epsilon$ and a specific kinetic energy $\frac{1}{2}\vB\cdot\vB$ contribution
and is given as 
\begin{equation}
	e=\epsilon+\frac{1}{2}\vB\cdot\vB.
	\label{eq:energy}
\end{equation}
Since the resulting equation system consists of $5+\left(N_k-1\right)$ equations for $5+N_k$ unknowns, a caloric and a thermal equation of state (EOS) are required 
for closure. They relate the density $\rho$, temperature $T$ and concentration $\YB$ to the specific internal energy $\epsilon$ and pressure $p$ respectively: 
\begin{equation}
	\epsilon=\epsilon(\rho,T,\YB),\quad p=p(\rho,T,\YB).
	\label{eq:eos_general}
\end{equation}
The numerical studies in this work are restricted to mixtures of ideal gases where the caloric and thermal EOS have the form
\begin{equation}
	\epsilon=\epsilon(\rho,T,\YB):=\frac{\mathcal{R}_{\text{mix}}}{\kappa_{\text{mix}}-1}T, \quad p=p(\rho,T,\YB):=\rho  T\mathcal{R}_{\text{mix}}. 
	\label{eq:eos}
\end{equation}
For a mixture of ideal gases, Dalton's law justifies the assumption of linear mixture rules. Therefore, the ideal gas constant $\mathcal{R}_{\text{mix}}$ 
and the ratio of heat capacities $\kappa_{\text{mix}}$ can be expressed as
\begin{equation}
	\mathcal{R}_{\text{mix}}=c_{p,\text{mix}}-c_{v,\text{mix}} ,\quad \kappa_{\text{mix}}=\frac{c_{p,\text{mix}}}{c_{v,\text{mix}}}
	\label{eq:mix_R_kappa}
\end{equation}
with
\begin{equation}
	c_{p,\text{mix}} = \sum_{j=1}^{N_k}Y_j\cdot c_{p,j},\quad c_{v,\text{mix}} = \sum_{j=1}^{N_k}Y_j\cdot c_{v,j}.
	\label{eq:mix_cp_cv}
\end{equation}
Linear mixture rules also apply for the dynamic viscosity $\mu_{\text{mix}}$ and the Prandtl number $\text{Pr}_{\text{mix}}$ 
\begin{equation}
	\mu:=\mu_{v,\text{mix}} = \sum_{j=1}^{N_k}Y_j\cdot \mu_{j},\quad \text{Pr}_{\text{mix}} := \sum_{j=1}^{N_k}Y_j \cdot \text{Pr}_{j}.
	\label{eq:mix_lambda_mu}
\end{equation}
Finally, the heat conductivity $\lambda_\text{mix}$ can be computed as
\begin{equation}
	\lambda_{\text{mix}}=\frac{\mu_{\text{mix}}c_{p,\text{mix}}}{\text{Pr}_{\text{mix}}}.
\end{equation}
Throughout this paper, the species viscosity $\mu_k$, heat conductivities $\lambda_k$ and diffusion coefficients $D_k$ are assumed as constant, 
if not stated otherwise. 

\section{Numerical Method}
\label{sec:Numerics}
The objective of this section is to establish an hp-adaptive hybrid discontinuous Galerkin method with \FV sub-cell shock capturing for the simulation
of compressible turbulent multi-component flows. While the hybrid hp-adaptive \DGFV operator was previously introduced for gas-dynamics in~\cite{Mossier2022} 
and applied for inviscid sharp-interface simulations in~\cite{Mossier2023}, the present paper extends this to viscous multi-component flows on non-conforming grids. Therefore,
the following section is organized as follows. First, the p-adaptive DGSEM operator and the h-refined \FV operator are revisited and extended by parabolic terms. 
Then, the coupling between the operators is addressed with an extension to non-conforming grids, to facilitate a more flexible grid generation. 
Subsequently, the indicator scheme is briefly discussed, accounting for the challenges of multi-component flows. Finally, a dealiasing strategy and a
positivity preserving filter are discussed. They pertain to the non-linear stability in the presence of under-resolved turbulence and complex shock-patterns.  

\subsection{hp-Adaptive Discretization}
\label{subsec:hp_operator}
The compressible Navier--Stokes Equations \eqref{eq:ns_flux_notation} are discretized on a computational domain $\Omega\subset \mathbb{R}^3$,
which is subdivided into $K\in\mathbb{N}$ non-overlapping hexahedral elements $\Omega^E$ such that $\Omega=\bigcup^K_{e=1}\Omega^E_e$ and $\bigcap^K_{e=1}\Omega^E_e=\emptyset$ holds.
While the tensor basis elements $\Omega^E$ are discretized with a discontinuous Galerkin spectral element method,
each \DG element $\Omega^E$ can be subdivided into a sub-grid of $N_{\text{FV}}$ \FV sub-cells $e^\Omega$ per direction, 
in accordance with a suitable indicator function as discussed in Section \ref{subsec:indicator}. 
On these \FV sub-cell elements, a second-order finite volume operator is applied. 
The basic idea is thus to advance an element via the DG operator for sufficiently smooth solutions and fall back to an
FV operator for non-smooth solution features. 
In the following, a short recap of the DGSEM and \FV operators for the Navier--Stokes equations 
is provided for general ansatz degrees $N$ and \FV sub-cell resolutions $N_{\text{FV}}$. For a detailed derivation of the DGSEM and \FV sub-cell operator,
the reader is referred to~\cite{Kopriva2009,Kaiser2020,Sonntag2014}.

\subsubsection{DGSEM Operator}
\label{subsec:p_DG_operator}
The DGSEM operator is derived for a reference element $E=[-1,1]^3$ that is linked to the physical element $\Omega^E$ through a 
mapping from physical coordinates $\xB=\left(x_1,x_2,x_3\right)^T$ to reference coordinates $\xiB=\left(\xi_1,\xi_2,\xi_3\right)^T$. 
In reference space, the Equation \eqref{eq:ns_flux_notation} reads 
\begin{equation}
	J_{\mathrm{geo}}\frac{\partial\uB}  {\partial t} + \nabla_{\xiB} \cdot \FtB(\uB,\nabla \uB) = 0, \label{eq:ns_trafo} 
\end{equation}
with the Jacobi determinant $J_{\mathrm{geo}}$ of the mapping and the contravariant flux $\FtB$. Since Equation \eqref{eq:ns_trafo} involves gradients of the solution,
we follow the \textit{lifting} strategy of Bassi and Rebay~\cite{Bassi1997} and rewrite \eqref{eq:ns_trafo} as a system of first-order equations:
\begin{subequations}
	\begin{align}
		J_{\mathrm{geo}}\frac{\partial\uB}  {\partial t} + \nabla_{\xiB} \cdot \FtB(\uB,\gB) &=0, \label{eq:ns_trafo_lift} \\
		\gB^d-\frac{1}{J_{\mathrm{geo}}} \nabla_{\xiB} \cdot  \boldsymbol{\mathcal{U}}^d &=0, \quad d=1,2,3\label{eq:lift_trafo} 
	\end{align}
\end{subequations}
with the contravariant solution in the direction $d$ denoted as $\boldsymbol{\mathcal{U}}^d$. By projecting \eqref{eq:ns_trafo_lift} and \eqref{eq:lift_trafo}
onto a space of polynomial test functions $\psi\in\mathbb{P}$ and integration by parts, a weak form of both equations is obtained as
\begin{subequations}
	\begin{align}
		\int_{E}(J_{\mathrm{geo}}\frac{\partial\uB}{\partial t})\psi d\Omega + \oint_{\partial E} (\FtB\cdot\nB_{\xiB}) \psi d S_\xi- \int_{E}(\FtB\cdot\nabla_{\xiB})\psi d\Omega &= 0, \label{eq:ns_weak}\\
		\int_{E}(J_{\mathrm{geo}}\gB^d)\psi d\Omega + \oint_{\partial E} (\boldsymbol{\mathcal{U}}^d\cdot\nB_{\xiB}) \psi d S_\xi- \int_{E}(\boldsymbol{\mathcal{U}}^d\cdot\nabla_{\xiB})\psi d\Omega &= 0.\label{eq:lift_weak}
	\end{align} 
\end{subequations}
Subsequently, a piece-wise polynomial ansatz is introduced for the solution $\uB$, the contravariant flux $\FtB$, the
lifting variable $\gB^d$ and the contravariant lifting flux $\boldsymbol{\mathcal{U}}^d$
\begin{align}
	\uB(\xiB,t) &\approx\sum^N_{i,j,k=0}\hat{\QB}_{ijk}(t)\zeta_{ijk}(\xiB),           &\, \gB^d(\xiB,t) &\approx\sum^N_{i,j,k=0}\hat{\gB}_{ijk}(t)\zeta_{ijk}(\xiB),\\
	\FtB(\xiB,t)&\approx\sum^N_{i,j,k=0}\hat{\FtB}_{ijk}(\hat{\QB}_{ijk})\zeta_{ijk}(\xiB),  &\, \boldsymbol{\mathcal{U}}^d(\xiB,t) &\approx\sum^N_{i,j,k=0}\hat{\boldsymbol{\mathcal{U}}}^d_{ijk}(t)\zeta_{ijk}(\xiB)
\end{align}
in a solution space spanned by tensor products of one-dimensional \textit{Lagrange} polynomials
\begin{equation}
	\zeta_{ijk}(\xiB) = \ell_i(\xi^1)\ell_j(\xi^2)\ell_k(\xi^3).
	\label{eq:langrange_basis}
\end{equation}
In line with the Galerkin approach, the Lagrange basis \eqref{eq:langrange_basis} is used for both the ansatz functions $\zeta$ and the test functions $\psi$.

Since the piece-wise polynomial ansatz allows for discontinuities between elements, 
the contravariant flux $\smash{(\FtB\cdot\nB_{\xiB})}$ and the contravariant solution $(\boldsymbol{\mathcal{U}}^d\cdot\nB_{\xiB})$
are not uniquely defined at element interfaces. While a simple arithmetic mean 
\begin{align}
	(\boldsymbol{\mathcal{U}}^d\cdot\nB_{\xiB})\approx \UB^{*} &= \frac{1}{2}(\uB^-+\uB^+),            \label{eq:lift_flux}\\
	(\FtB^v\cdot\nB_{\xiB})\approx \FB^{v,*} &= \frac{1}{2}(\FtB^v(\uB^-,\gB^-)+\FtB^v(\uB^+,\gB^+)) , \label{eq:visc_flux} 
\end{align}
can be assumed for the inter-element solution in the lifting Equation \eqref{eq:lift_weak} and the viscous part of the contravariant flux 
in the main Equation \eqref{eq:ns_weak}, the convective flux is replaced by a characteristics based numerical flux
\begin{equation}
	(\FtB^c\cdot\nB_{\xiB})\approx \FB^{c,*}\left(\uB^-,\uB^+,\nB_{\xiB}\right) \label{eq:conv_flux} 
\end{equation}
which is computed by a Riemann solver. Here, the superscripts $^-$ and $^+$ indicate the surface solution from both sides at an element interface. 
The present study employs the Roe solver~\cite{Roe1981} with the entropy fix of Harten and Hyman~\cite{Harten1983} as an approximate Riemann solver
when not stated otherwise. 

To obtain a semi-discrete scheme, the integrals in Equations \eqref{eq:ns_weak} and \eqref{eq:lift_weak} are replaced
by a numerical quadrature. A defining feature of the DGSEM is the \textit{collocation} of interpolation and quadrature nodes,
which establishes a tensor product structure of the DGSEM operator and reduces the computational cost per degree of freedom (DOF).
In this work, either \textit{Legendre--Gauss} (LG) nodes or \textit{Legendre--Gauss--Lobatto} (LGL) nodes are chosen. 
Owing to the tensor product structure, a multi-dimensional DGSEM discretization is given as a succession of one-dimensional 
operations. The semi-discrete form of \eqref{eq:ns_weak} and \eqref{eq:lift_weak} is finally obtained as  
\begin{align}
	\frac{\partial\hat{\uB}_{ijk}}{\partial t} = \frac{-1}{\JB_{ijk}} \bigg[ 
		& \sum^N_{m=0} \hat{\FtB}_{mjk}^1\hat{\DB}_{im} +
		\left[\FB^*\hat{s}\right]_{jk}^{+\xi_1} \hat{\ell}_i(1) +
		\left[\FB^*\hat{s}\right]_{jk}^{-\xi_1} \hat{\ell}_i(-1) \notag \\
		& + \sum^N_{n=0} \hat{\FtB}_{ink}^2\hat{\DB}_{jn} +
		\left[\FB^*\hat{s}\right]_{ik}^{+\xi_2} \hat{\ell}_j(1) +
		\left[\FB^*\hat{s}\right]_{ik}^{-\xi_2} \hat{\ell}_j(-1) \notag \\
		& + \sum^N_{o=0} \hat{\FtB}_{ijo}^3\hat{\DB}_{ko} +
		\left[\FB^*\hat{s}\right]_{ij}^{+\xi_3} \hat{\ell}_k(1) +
		\left[\FB^*\hat{s}\right]_{ij}^{-\xi_3} \hat{\ell}_k(-1)
	\bigg]
	\label{eq:semidiscrete}
\end{align}
and
\begin{align}
	\hat{\gB}^d_{ijk} = \frac{-1}{\JB_{ijk}} \bigg[ 
		& \sum^N_{m=0} \hat{\boldsymbol{\mathcal{U}}}_{mjk}^{1,d}\hat{\DB}_{im} +
		\left[\UB^{*,d}\hat{s}\right]_{jk}^{+\xi_1} \hat{\ell}_i(1) +
		\left[\UB^{*,d}\hat{s}\right]_{jk}^{-\xi_1} \hat{\ell}_i(-1) \notag \\
		& + \sum^N_{n=0} \hat{\boldsymbol{\mathcal{U}}}_{ink}^{2,d}\hat{\DB}_{jn} +
		\left[\UB^{*,d}\hat{s}\right]_{ik}^{+\xi_2} \hat{\ell}_j(1) +
		\left[\UB^{*,d}\hat{s}\right]_{ik}^{-\xi_2} \hat{\ell}_j(-1) \notag \\
		& + \sum^N_{o=0} \hat{\boldsymbol{\mathcal{U}}}_{ijo}^{3,d}\hat{\DB}_{ko} +
		\left[\UB^{*,d}\hat{s}\right]_{ij}^{+\xi_3} \hat{\ell}_k(1) +
		\left[\UB^{*,d}\hat{s}\right]_{ij}^{-\xi_3} \hat{\ell}_k(-1)
	\bigg]
	\label{eq:semidiscrete_lift}
\end{align}
respectively, with the abbreviations
\begin{equation}
    \hat{\ell}_i=\frac{\ell_i}{\omega_i},\quad \hat{\DB}_{ij}=-\frac{\omega_i}{\omega_j}\DB_{ji},\quad \DB_{ij}=\frac{\partial\ell_j(\xi)}{\partial\xi}\bigg\rvert_{\xi=\zeta_i},
    \label{eq:dgsem_abbrev}
\end{equation}
denoting the weighted Lagrange polynomials $\hat{\ell}_i$, an entry of the weighted differentiation matrix $\hat{\DB}_{ij}$
and the derivative of a Lagrange basis polynomial $\DB_{ij}$ respectively.

\subsubsection{FV Sub-Cell Operator}
\label{subsec:h_FV_operator}
The DGSEM's piece-wise polynomial solution is highly accurate in smooth regions but notoriously unstable for elements containing 
discontinuous solution features like shocks, material interfaces or under-resolved gradients due to aliasing. 
As a stabilization technique, Huerta et al.~\cite{Huerta2012} and Persson et al.~\cite{Persson2013} 
proposed to include piece-wise constant functions in the ansatz and test spaces. 
To alleviate the accuracy loss of the low-order ansatz, the \FV method is applied on a 
sub-grid in each \DG element $\Omega^E$ with $N_\text{FV}$ sub-cells
per dimension $d$. In reference space, the resulting equidistant grid consists of $N_\text{FV}^d$ sub-cells $e^\Omega_{ijk}$
with a characteristic length of $\smash{l_{\text{FV}}=\frac{2}{N_\text{FV}}}$. Assuming a piece-wise constant basis,  
Equation \eqref{eq:ns_weak} reduces to
\begin{subequations}
	\begin{align}
		\int_{e^\Omega_{ijk}}(J^{\text{FV}}_{\mathrm{geo}}\frac{\partial\uB}{\partial t}) d\Omega + \oint_{\partial e^\Omega_{ijk}} (\FtB\cdot\nB_{\xiB}) d S_\xi &= 0,\label{eq:ns_weak_fv}
	\end{align} 
\end{subequations}
where $J^{\text{FV}}_{\mathrm{geo}}$ denotes the Jacobian of the mapping to a reference sub-cell $e^\Omega_{ijk}$. Subsequently,
integral mean values $\hat{\uB}^\text{FV}_{ijk}$ are introduced for the solution and flux integrals are replaced by the midpoint rule. 
The resulting evolution equation for the integral mean values reads as
\begin{equation}
    \begin{split}
        \frac{\hat{\uB}^\text{FV}_{ijk}}{\partial t}=-\frac{1}{J^{\text{FV}}_{ijk}}\bigg[
        &\left[\FB^*\hat{s}\right]_{i-\frac{1}{2}jk}+\left[\FB^*\hat{s}\right]_{i+\frac{1}{2}jk} \\
        &\left[\FB^*\hat{s}\right]_{ij-\frac{1}{2}k}+\left[\FB^*\hat{s}\right]_{ij+\frac{1}{2}k} \\
        &\left[\FB^*\hat{s}\right]_{ijk-\frac{1}{2}}+\left[\FB^*\hat{s}\right]_{ijk+\frac{1}{2}} \bigg].
        \label{eq:semidiscrete_fv}
    \end{split}
\end{equation}
As for the \DG operator, the non-unique surface terms $\smash{(\FtB^c\cdot\nB_{\xiB})}$ of the convective flux contribution 
are replaced with a numerical flux $\smash{(\FtB^c\cdot\nB_{\xiB})}\approx \FtB^{c,*}(\uB^-,\uB^+,\nB_{\xiB})$, provided by a Riemann solver.
In the context of the \FV scheme, the second-order derivatives in the viscous flux can be calculated as central differences. 

The accuracy of the \FV scheme can be further enhanced with a second-order TVD reconstruction
as proposed by Sonntag and Munz~\cite{Sonntag2014}. Throughout this work, we employ a \textit{minmod} function to limit the reconstruction.  

\subsection{Assembly of hp-adaptive Operator}
\label{subsec:coupling}
For the construction of an hp-adaptive hybrid \DGFV scheme, we allow for a variable element-local degree $N\in[N_\text{min},N_\text{max}]$ 
and an \FV sub-cell resolution $N_\text{FV}$, which can be chosen independently of the \DG ansatz $N$.
The bounds $N_\text{min}$ and $N_\text{max}$ as well as the \FV resolution are problem dependent and have to be selected prior to a computation,
to precompute all required geometrical metrics and mappings between the operators. 
Based on a comparison of the explicit time step conditions for the DG and FV operators, Dumbser et al.~\cite{Dumbser2016a} proposed 
a sub-cell resolution of up to $N_\text{FV} = 2N_\text{max}+1$ to maximize resolution without impeding the overall timestep constraint.
To assemble the dynamic hp-adaptive method, three key aspects have yet to be addressed: 
\begin{itemize}
	\item To achieve dynamic refinement at runtime, the solution of a \DG element has to be transformed between different ansatz degrees $N\neq \tilde{N}$
	with the mappings $\NTOM$ and $\MTON$.
	\item \FV shock capturing necessitates the switching between a piece-wise polynomial \DG and the piece-wise constant \FV representation
	with the mappings $\FVDGB$ and $\DGFVB$.
	\item In the presence of variable ansatz degrees and both \FV and \DG elements, the computation of convective, viscous and lifting fluxes needs 
	to account for non-conforming solution representations at element interfaces.
\end{itemize}

The mappings $\NTOM$ for $N < \tilde{N}$ can be achieved exactly through an interpolation, while $\MTON$ for $N > \tilde{N}$ requires a projection
to the lower degree. Mapping a polynomial solution to piece-wise constant sub-cell data, $\DGFVB$, is achieved via integration over each sub-cell. 
The inverse mapping $\FVDGB$ is computed as the pseudo-inverse of $\DGFVB$, following Dumbser et al.~\cite{Dumbser2016a}. A detailed derivation 
of these mappings is performed in~\cite{Sonntag2014,Mossier2022}.

In the following, the numerical flux computation is described for both convective and viscous contributions $F^*=F^{c,*}+F^{v,*}$ at element interfaces with mixed discretizations. 
Particular attention is devoted to the case of non-conforming computational grids.

\subsubsection{Coupling of Convective Terms}
\label{subsubsec:conv_coupling}
The convective component of the numerical flux $F^{c,*}$ is determined by solving a Riemann problem. This necessitates a common representation of 
the solution at an element interface. As previously stated in~\cite{Mossier2022}, this involves the transformation to a common ansatz degree
$N=\text{max}(N,\tilde{N})$ with $\NTOM$ in case of a \DG interface with variable ansatz degree.

\begin{figure}[t]
	\centering
	\includegraphics[width=0.8\textwidth]{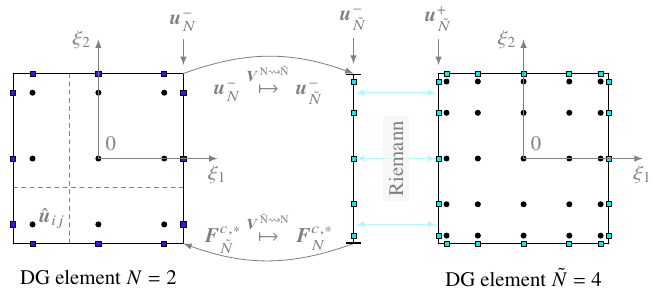}
	\caption{Computation of the convective flux $\FB^{c,*}$ at element interfaces with mixed ansatz degree $N\neq\tilde{N}$. 
	The surface solution $\uB^-_N$ of the left DG element is interpolated from $N=2$ to $\tilde{N}=4$ and stored in $\uB^-_{\tilde{N}}$. 
	Subsequently, the convective numerical flux $F^{c,*}(\uB^-_{\tilde{N}},\uB^+_{\tilde{N}})$ is computed on the common ansatz degree $\tilde{N}=4$.
	Finally, the flux $F^{c,*}$ is projected to the native degree $N=2$ of the left element.}
	\label{fig:p_coupling}
\end{figure}
\begin{figure}[h!]
	\centering
	\includegraphics[width=0.8\textwidth]{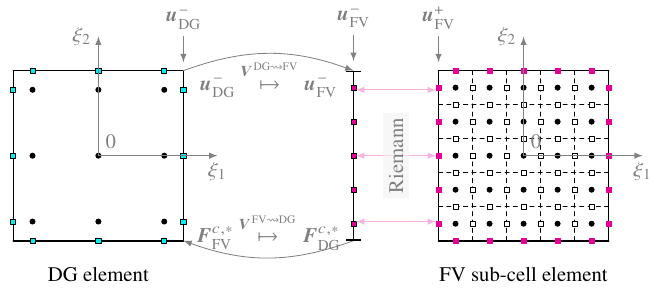}
	\caption{Computation of the convective flux $\FB^{c,*}$ at mixed \DGFV element interfaces.
	The surface solution $\uB^-_{\text{DG}}$ of the left element is projected from a polynomial representation 
	to a piece-wise constant FV representation and stored in $\uB^-_{\text{FV}}$. 
	Subsequently, the convective numerical flux $F^{c,*}(\uB^-_{\text{FV}},\uB^+_{\text{FV}})$ is computed with the piece-wise constant data
	and finally transformed back to the native polynomial representation of the left element.}
	\label{fig:h_coupling}
\end{figure}

At mixed \DGFV interfaces, the flux computation calls for 
the transformation of the \DG side solution to an \FV representation $\DGFVB$ to evaluate the flux on the piece-wise constant \FV data. 
Once the numerical flux \eqref{eq:conv_flux} is computed, it is transformed back into the native discretization of each element for application in the surface integral
via $\MTON$ or $\FVDGB$ respectively. Figures \ref{fig:p_coupling} and \ref{fig:h_coupling} illustrate the flux computation for both types of mixed interfaces.

\subsubsection{Coupling of Viscous Terms}
\label{subsubsec:visc_coupling}
In this work, the hybrid hp-adaptive \DGFV discretization is applied to the full compressible Navier--Stokes equations for multi-component flows for the first time. 
This involves discretizing viscous fluxes related to shear-stresses \eqref{eq:visc_stress}, heat conduction \eqref{eq:fourier_law} and species diffusion \eqref{eq:fickian_law}
and the solution of an additional lifting Equation \eqref{eq:ns_trafo_lift} to obtain second derivatives.

\begin{figure}[t]
	\centering
	\includegraphics[width=0.8\textwidth]{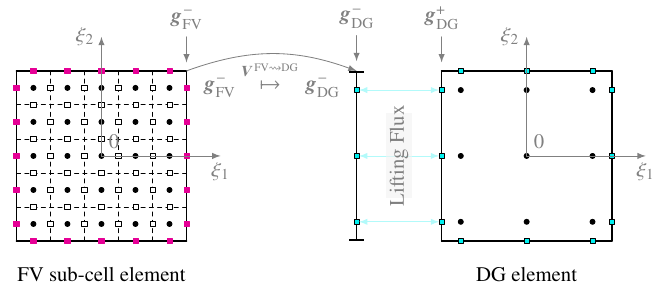}
	\caption{Computation of the lifting flux $\UB^*$ at mixed \DGFV element interfaces.	The surface solution $\gB^-_{\text{FV}}$ of the 
	left element is projected a piece-wise constant FV representation to a polynomial representation and stored in $\gB^-_{\text{DG}}$. 
	Subsequently, the lifting flux $\UB^{*}(\gB^-_{\text{DG}},\gB^+_{\text{DG}})$ is computed with the polynomial data for the DG element.}
	\label{fig:h_coupling_lift}
\end{figure}
\begin{figure}[h!]
	\centering
	\includegraphics[width=0.8\textwidth]{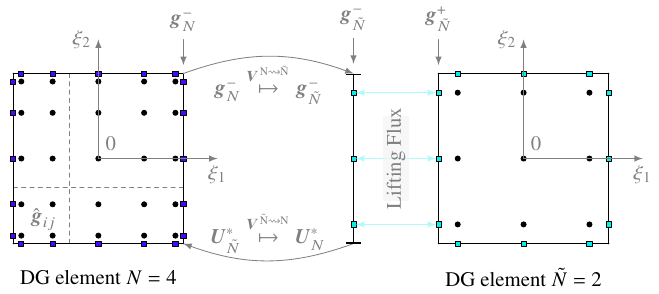}
	\caption{Computation of the lifting flux $\UB^*$ at element interfaces with mixed ansatz degree $N\neq\tilde{N}$.
	The surface solution $\gB^-_N$ of the left DG element is projected from $N=4$ to $\tilde{N}=2$ and stored in $\gB^-_{\tilde{N}}$. 
	Subsequently, the lifting flux $\UB^{*}(\gB^-_{\tilde{N}},\gB^+_{\tilde{N}})$ is computed on the common ansatz degree $\tilde{N}=2$.
	Finally, the flux $\UB^{*}$ is interpolated to the native degree $N=4$ of the left element.}
	\label{fig:p_coupling_lift}
\end{figure}

The viscous fluxes $F^{v,*}$ are evaluated analogously to the convective fluxes $F^{c,*}$ on a common ansatz degree $N=\text{max}(N,\tilde{N})$ 
or on piece-wise constant \FV data in the case of mixed \DGFV interfaces. The sole difference to the convective fluxes lies in the flux computation itself.
Here, a simple arithmetic mean \eqref{eq:visc_flux} proves sufficient for the parabolic terms. 

Treatment of the lifting fluxes $\UB^*$ at mixed interfaces calls for a slightly different procedure. With lifting solely performed for the \DG discretization,
$\UB^*$ is evaluated on a pure \DG representation at mixed \DGFV interfaces by transforming the \FV side solution with $\FVDGB$ to polynomial data as
depicted in Figure \ref{fig:h_coupling_lift}. 
When non-conforming ansatz degrees $N\neq\tilde{N}$ are present, the lifting fluxes are computed on the lower degree $N=\text{min}(N,\tilde{N})$, as
shown in Figure \ref{fig:p_coupling_lift}.   

\subsubsection{Non-Conforming Interfaces}
\label{subsubsec:mortar_coupling}

Mesh generation with hexahedral elements remains challenging, especially for complex geometries. Allowing for non-conforming computational grids 
alleviates this to some extent and provides an efficient way to increase the spatial resolution locally. 
Treatment of non-conforming grids was addressed with the Mortar Element Method (MEM) 
by Maday et al~\cite{Maday1989} and Mavriplis et al.~\cite{Mavriplis1989,Feng2005}, and further analyzed by 
Bernardi et al.~\cite{Bernardi1990,Bernardi1994}. The \textit{mortar} technique was later applied to DGSEM by Kopriva~\cite{Kopriva2001} 
and Chalmers et al.~\cite{Chalmers2019} and adapted to mixed DG/FV discretizations by Sonntag~\cite{SonntagPhD} and Krais et al.~\cite{flexi}.
With this paragraph, we generalize the mortar technique to cope with both p-refinement and arbitrary \FV sub-cell resolutions.
\begin{figure}[t]
	\centering
	\begin{subfigure}[t]{0.49\textwidth}
		\includegraphics[width=1.0\textwidth]{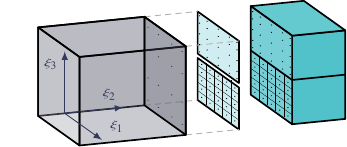}
		\caption{Mortar Type 1 and 2}
		\label{subfig:subfig:mortar_12}
	\end{subfigure}
	\hfill
	\begin{subfigure}[t]{0.49\textwidth}
		\includegraphics[width=1.0\textwidth]{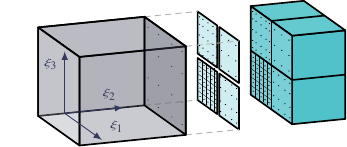}
		\caption{Mortar Type 3}
		\label{subfig:mortar_3}
	\end{subfigure}
	\caption{Schematic of the supported mortar types with Types 1 and 2 differing only in their orientation to the reference direction. 
	The image illustrates the mapping of the big mortar side to \textit{virtual} small mortar sides, that allow a flux computation on conforming surface data.}
	\label{fig:mortarTypes}
\end{figure}

Three different non-conforming interfaces are supported and illustrated in Figure \ref{fig:mortarTypes}. 
Types $1$ and $2$ consist of a big side connected to two small mortar sides, each of which is exactly half of the big side in reference space.
While Type $1$ and $2$ simply refine the small sides in different reference directions, Type $3$ can be interpreted as the combination of both and
provides a two to one refinement in both directions. Therefore, we can discuss the mortar technique based on Type $1$ interfaces without loss of generality.

\begin{figure}[t]
	\centering
	\tikzsetnextfilename{mortar_big_fv}
	\begin{subfigure}[t]{0.49\textwidth}
   \begin{tikzpicture}[x=2cm,y=2cm,>=triangle 45,line join=round, scale=1.0]
         \draw[thick] (-1,-1) node[anchor=east,inner sep=8pt] {\footnotesize$\uBZ^+$} -- (1,-1) node[anchor=west,inner sep=8pt] {\footnotesize$\uBZ^+$} -- (1,0) -- (-1,0) node[anchor=north west] {\footnotesize FV} -- cycle;
         \foreach \x in \xFV {
            \draw[fill=fvcolor] (\x,-1) +(-0.02,-0.02) rectangle +(0.02,0.02);
         }
         \foreach \a in \innerborders {
            \pgfmathsetmacro{\xx}{\a}
            \pgfmathsetmacro{\yy}{\a/2-0.5}
            \draw [dash pattern=on 2 off 2] (\xx, -1) -- (\xx, 0); 
            \draw [dash pattern=on 2 off 2] (-1, \yy) -- (1, \yy); 
         }
         
         \draw[thick] (-1,-3.25)node[fill=white,anchor=south west] {\footnotesize FV}  -- (0,-3.25) -- (0,-2.25) -- (-1,-2.25) node[anchor=east,inner sep=8pt] {\footnotesize$\uL^-$} -- cycle;
         \draw[thick,|-|] (-1,-1.5)  -- (0,-1.5);
         \draw[-latex] (-1,-1) to [bend right=45] node[anchor=west] {\tiny $\MB^{\mathrm{\footnotesize FV}}_{01} \cdot$} (-1,-1.5) node[anchor=east,inner sep=8pt] {\footnotesize$\uL^+$} ;
         \foreach \x in \xFV {
            \pgfmathsetmacro{\xx}{\x/2-0.5}
            \draw[fill=fvcolor] (\xx,-1.5) +(-0.02,-0.02) rectangle +(0.02,0.02);
         }
         \foreach \x in \xFV {
            \pgfmathsetmacro{\xx}{\x/2-0.5}
            \draw[fill=fvcolor] (\xx,-2.25) +(-0.02,-0.02) rectangle +(0.02,0.02);
            \draw[latex-latex,line width=0.007mm,fvcolor] (\xx,-1.5) -- (\xx,-2.25);
         }
         
         \foreach \a in \innerborders {
            \pgfmathsetmacro{\xx}{\a/2-0.5}
            \pgfmathsetmacro{\yy}{\a/2-2.75}
            \draw [dash pattern=on 2 off 2] (\xx, -2.25) -- (\xx, -3.25); 
            \draw [dash pattern=on 2 off 2] (-1, \yy) -- (0, \yy); 
         }
      
         \fill[fill=gray!10,opacity=0.6] (-0.5,-1.85)  +(-0.4,-0.12) rectangle +(0.4,0.10);
         \node[text=gray] at (-0.5,-1.85)  {\rotatebox{0}{\footnotesize Riemann}};
      
         \draw[thick,|-|] (0,-1.5) -- (1,-1.5) ;
         \draw[-latex] (1,-1) to [bend left=22.5] node[anchor=east] {\tiny$\MB^{\mathrm{\footnotesize FV}}_{02} \cdot$} (1,-1.5) node[anchor=west,inner sep=8pt] {\footnotesize$\uR^+$};
         \foreach \x in \xFV {
            \pgfmathsetmacro{\xx}{\x/2+0.5}
            \draw[fill=fvcolor] (\xx,-1.5) +(-0.02,-0.02) rectangle +(0.02,0.02);
            \draw[latex-latex,line width=0.007mm,fvcolor] (\xx,-1.5) -- (\xx,-1.9);
         }
         \draw[thick,|-|] (0,-1.9) -- (1,-1.9) ;
         \draw[-latex] (1,-2.25) to [bend right=22.5] node[anchor=east] {\tiny$\DGFVB \cdot$} (1,-1.9) node[anchor=west,inner sep=8pt] {\footnotesize$\uR^-$};
         \foreach \x in \xFV {
            \pgfmathsetmacro{\xx}{\x/2+0.5}
            \draw[fill=fvcolor] (\xx,-1.9) +(-0.02,-0.02) rectangle +(0.02,0.02);
         }
         \draw[thick] (0,-3.25) -- (1,-3.25) node[anchor=south east] {\footnotesize DG} -- (1,-2.25) node[anchor=west,inner sep=7pt] {\footnotesize$\uR^-$}  -- (0,-2.25) -- cycle;
         \foreach \x in \xGP {
            \pgfmathsetmacro{\xx}{\x/2+0.5}
            \draw[fill=dgcolor] (\xx,-2.25) +(-0.02,-0.02) rectangle +(0.02,0.02);
         }
         \fill[fill=gray!10,opacity=0.6] (0.5,-1.7)  +(-0.4,-0.10) rectangle +(0.4,0.10);
         \node[text=gray] at (0.5,-1.7)  {\rotatebox{0}{\footnotesize Riemann}};
         \foreach \x in \xGP {
         \foreach \y in \xGP {
         \draw[fill=gray] (\x/2+0.5,\y/2-2.75) circle (0.02);
         }
         }
   \end{tikzpicture}
   \caption{}
   \label{fig:mortar_setup_1}
\end{subfigure}
\hfill
\begin{subfigure}[t]{0.49\textwidth}
   \begin{tikzpicture}[x=2cm,y=2cm,>=triangle 45,line join=round, scale=1.0]
      \draw[thick] (-1,-1) node[anchor=east,inner sep=8pt] {\footnotesize$\uBZ^+$} -- (1,-1) node[anchor=west,inner sep=8pt] {\footnotesize$\uBZ^+$} -- (1,0) -- (-1,0) node[anchor=north west] {\footnotesize DG} -- cycle;
      \foreach \x in \xGPI {
         \draw[fill=dgcolor] (\x,-1) +(-0.02,-0.02) rectangle +(0.02,0.02);
      }
      \foreach \x in \xGPI {
         \foreach \y in \xGPI {
         \draw[fill=gray] (\x,\y*0.5-0.5) circle (0.02);
         }
      }
      \draw[thick] (-1,-3.25) node[fill=white,anchor=south west] {\footnotesize FV} -- (0,-3.25) -- (0,-2.25) -- (-1,-2.25) node[anchor=east,inner sep=8pt] {\footnotesize$\uL^-$} -- cycle;
      \draw[thick,|-|] (-1,-1.5)  -- (0,-1.5);
      \draw[-latex] (-1,-1) to [bend right=45] node[anchor=west] {\tiny$\MB_{01} \cdot$} (-1,-1.5);
      \foreach \x in \xGPI {
         \pgfmathsetmacro{\xx}{\x/2-0.5}
         \draw[fill=dgcolor] (\xx,-1.5) +(-0.02,-0.02) rectangle +(0.02,0.02);
      }
      \draw[thick,|-|] (-1,-1.85) -- (0,-1.85);
      \draw[-latex] (-1,-1.5) to [bend right=45] node[anchor=west] {\tiny$\DGFVB $} (-1,-1.85) node[anchor=east,inner sep=8pt] {\footnotesize$\uL^+$} ;
      \foreach \x in \xFV {
         \pgfmathsetmacro{\xx}{\x/2-0.5}
         \draw[fill=fvcolor] (\xx,-1.85) +(-0.02,-0.02) rectangle +(0.02,0.02);
      }
      \foreach \x in \xFV {
         \pgfmathsetmacro{\xx}{\x/2-0.5}
         \draw[fill=fvcolor] (\xx,-2.25) +(-0.02,-0.02) rectangle +(0.02,0.02);
         \draw[latex-latex,line width=0.007mm,fvcolor] (\xx,-1.85) -- (\xx,-2.25);
      }
      \foreach \a in \innerborders {
         \pgfmathsetmacro{\xx}{\a/2-0.5}
         \pgfmathsetmacro{\yy}{\a/2-2.75}
         \draw [dash pattern=on 2 off 2] (\xx, -2.25) -- (\xx, -3.25); 
         \draw [dash pattern=on 2 off 2] (-1, \yy) -- (0, \yy); 
      }
      \fill[fill=gray!10,opacity=0.6] (-0.5,-2.05)  +(-0.4,-0.12) rectangle +(0.4,0.10);
      \node[text=gray] at (-0.5,-2.05)  {\rotatebox{0}{\footnotesize Riemann}};

      \draw[thick,|-|] (0,-1.5) -- (1,-1.5) ;
      \draw[-latex] (1,-1) to [bend left=22.5] node[anchor=east] {\tiny $\MB_{02} \cdot$} (1,-1.5) node[anchor=west,inner sep=8pt] {\footnotesize$\uR^+$};
      \foreach \x in \xGPI {
         \pgfmathsetmacro{\xx}{\x/2+0.5}
         \draw[fill=dgcolor] (\xx,-1.5) +(-0.02,-0.02) rectangle +(0.02,0.02);
      }
      \draw[thick,|-|] (0,-1.85) -- (1,-1.85) ;
      \draw[-latex] (1,-1.5) to [bend right=-22.5] node[anchor=east] {\tiny $\NTOM \cdot$} (1,-1.85) node[anchor=west,inner sep=8pt] {\footnotesize$\uR^+$};
      \foreach \x in \xGP {
         \pgfmathsetmacro{\xx}{\x/2+0.5}
         \draw[fill=dgcolor] (\xx,-1.85) +(-0.02,-0.02) rectangle +(0.02,0.02);
         \draw[latex-latex,line width=0.007mm,dgcolor] (\xx,-2.25) -- (\xx,-1.85);
      }
      \draw[thick] (0,-3.25) -- (1,-3.25) node[anchor=south east] {\footnotesize DG} -- (1,-2.25) node[anchor=west,inner sep=7pt] {\footnotesize$\uR^-$}  -- (0,-2.25) -- cycle;
      \foreach \x in \xGP {
         \pgfmathsetmacro{\xx}{\x/2+0.5}
         \draw[fill=dgcolor] (\xx,-2.25) +(-0.02,-0.02) rectangle +(0.02,0.02);
      }
      \fill[fill=gray!10,opacity=0.6] (0.5,-2.05)  +(-0.4,-0.12) rectangle +(0.4,0.10);
      \node[text=gray] at (0.5,-2.05)  {\rotatebox{0}{\footnotesize Riemann}};
      \foreach \x in \xGP {
      \foreach \y in \xGP {
      \draw[fill=gray] (\x/2+0.5,\y/2-2.75) circle (0.02);
      }
   }
   \end{tikzpicture}
   \caption{}
   \label{fig:mortar_setup_2}
\end{subfigure}
	\caption{Flux computation at a non-conforming element interface with mixed hp-discretizations. 
	Setup \ref{fig:mortar_setup_1} shows a mortar interface where the larger element uses the \FV sub-cell discretization, 
	while Setup \ref{fig:mortar_setup_2} depicts a case with a \DG discretization for the larger element.}
	\label{fig:mortar_coupling}
\end{figure}
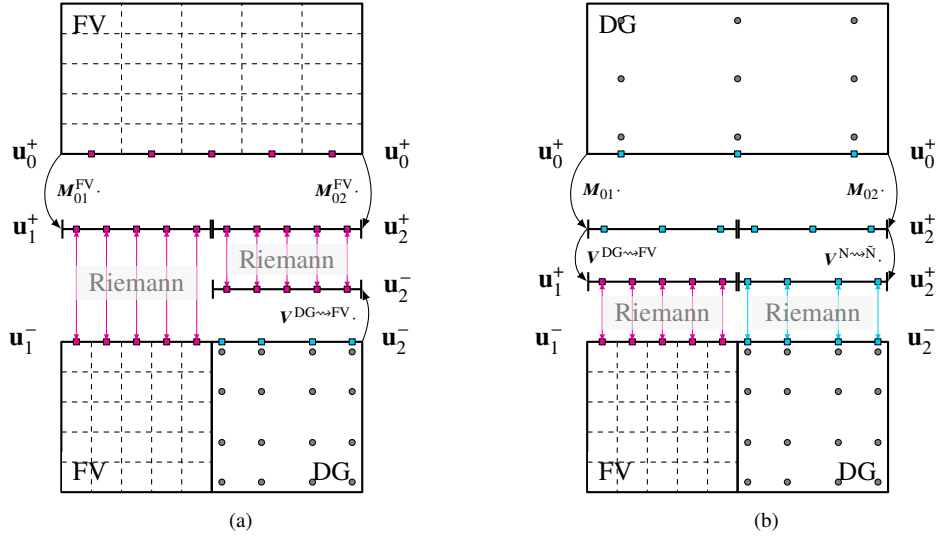

With the hp-adaptive operator, a panoply of different discretizations can be combined at a single non-conforming 
interface. To illustrate the flux computation at such an interface, Figure \ref{fig:mortar_coupling} depicts two possible scenarios 
involving different ansatz degrees and \FV sub-cell elements. 
The flux computation between non-conforming elements of different discretizations requires in general five key steps:

\begin{enumerate}
	\item The surface solution of the big mortar side $\uB_0^+$ is interpolated to virtual small sides $\uB_1^+$ and $\uB_2^+$ with 
	the Vandermonde matrices $\MB_{01}$ and $\MB_{02}$. When the surface solution uses a piece-wise constant \FV sub-cell representation, the mappings $\MB_{01}^\text{FV}$ and 
	$\MB_{02}^\text{FV}$ are used. The matrices can be precomputed at the start of a computation and are derived 
	in~\cite{SonntagPhD,flexi}. 
	\item In the presence of variable discretizations, the solutions $\uB_1^+$ and $\uB_2^+$ of the virtual small sides and the solution
	of the small sides $\uB_1^-$ and $\uB_2^-$ have to be transformed to a common solution representation. This is done with the 
	mappings $\DGFVB$, $\FVDGB$, $\MTON$ and $\NTOM$ according to the procedure described in Sections \ref{subsubsec:conv_coupling} and \ref{subsubsec:visc_coupling}.
	\item With surface data available in a conforming representation, the flux $\FB^*=\FB^{v,*}+\FB^{c,*}$ can be computed.
	\item Before applying the flux in the surface integral, it has to be transformed back to its original discretization using $\DGFVB$, $\FVDGB$, $\MTON$ or $\NTOM$.
	\item Finally, the flux is projected from the small virtual mortar sides to the big mortar side with $\FB^*_0=\MB_{10} \cdot \FB^*_1+ \MB_{20}\cdot \FB^*_2$
	or $\FB^*_0=\MB_{10}^\text{FV} \cdot \FB^*_1+ \MB_{20}^\text{FV}\cdot \FB^*_2$ when \FV sub-cell elements are involved. 
\end{enumerate}

\subsection{Temporal Discretization}
\label{subsubsec:timedisc}

To obtain a fully discrete scheme, the \DGFV operator is integrated in time with an explicit fourth-order low-storage Runge--Kutta (RK) scheme of Kennedy and 
Carpenter~\cite{Kennedy2003}. The explicit RK scheme is subject to the Courant Friedrichs Lewy (CFL) condition, which imposes 
time step constraints $\Delta t^c$ and $\Delta t^v$ related to the maximum hyperbolic signal velocities $\lambda^c$ and parabolic eigenvalues
$\delta^v$. For DG elements of degree $N$ and \FV sub-cell elements they are given as 
\begin{align}
	\Delta t^c_{\text{DG}} &= \text{CFL}^c\cdot\alpha_{\text{RK}}\left(N\right)\frac{\Delta x_\text{DG}}{\lambda^c\left(2N+1\right)},            \\
	\Delta t^c_{\text{FV}} &= \text{CFL}^c\cdot\alpha_{\text{RK}}\left(0\right)\frac{\Delta x_\text{FV}}{\lambda^c},   \quad \text{CFL}^c\in(0,1], \\
	\Delta t^v_{\text{DG}} &= \text{CFL}^v\cdot\beta_{\text{RK}}\left(N\right)\frac{\left(\Delta x_\text{DG}\right)^2}{\delta^v\left(2N+1\right)},           \\
	\Delta t^v_{\text{FV}} &= \text{CFL}^v\cdot\beta_{\text{RK}}\left(0\right)\frac{\left(\Delta x_\text{FV}\right)^2}{\delta^v}, \quad \text{CFL}^v\in(0,1], 
\end{align}
with the \DG and \FV sub-cell grid spacings $\Delta x_\text{DG}$ and $\Delta x_\text{FV}$ and the empirical scaling factors for the stability region 
$\alpha_{\text{RK}}$ and $\beta_{\text{RK}}$, which depend on the local ansatz degree. The coupled time step for the hyperbolic-parabolic system 
is finally obtained in terms of $\Delta t^c$ and $\Delta t^v$ as
\begin{equation}
	\Delta t=\frac{1}{\frac{1}{\underset{\Omega}{\text{min}}\left\{\Delta t^c\right\}}+\frac{1}{\underset{\Omega}{\text{min}}\left\{\Delta t^v\right\}}}.
\end{equation}

\subsection{Indicator Strategy}
\label{subsec:indicator}
Control over both p-adaptation and \FV sub-cell shock capturing is provided by an error estimator that analyses the decay rate of 
the modal polynomial solution representation~\cite{Mavriplis1989, Mavriplis1990, Mossier2022}. 
For a given element $E$, the element-local solution $\uB(\xiB)$ can be represented as an infinite series of polynomial 
basis functions $\zeta_{ijk}(\boldsymbol{\xi})$
\begin{equation}
    \uB(\boldsymbol{\xi})=\sum_{i,j,k=0}^{\infty} \hat{\uB}_{ijk}\zeta_{ijk}(\boldsymbol{\xi})=
   \underbrace{\sum_{i,j,k=0}^{N} \hat{\uB}_{ijk}\zeta_{ijk}(\boldsymbol{\xi})}_{\text{Ansatz}}+
   \underbrace{\sum_{i,j,k=N+1}^{\infty} \hat{\uB}_{ijk}\zeta_{ijk}(\boldsymbol{\xi})}_{\text{Truncation error}}.
    \label{eq:solution_expansion}
\end{equation}
It is evident from Equation \eqref{eq:solution_expansion} that an ansatz up to degree $N$ is associated with a truncation error. 
When transforming \eqref{eq:solution_expansion} to a modal Legendre basis with the Vandermonde matrix $\VB_{\text{Leg}}$,
the new coefficients $\hat{\uB}_{\text{Leg},ijk}$ correspond to the amplitudes of the respective solution modes.
The idea of the indicator is to infer an error estimate from the decay rate of these solution modes. 
Given a smooth solution, an exponential decay of the solution modes can be expected, while the extrapolation of the decay rate correlates with
the truncation error. Oscillatory solution behavior manifests as large amplitude of the highest solution modes, which 
leads to a very slow decay or even an increased contribution of the higher modes. 
The contribution $w_m^{\xi_i}$ of the $m^{\text{th}}$ solution mode in $\xi_i$-direction can be evaluated as
\begin{equation}
    w_{m}^{\xi_i}=\left[\frac{\left(\sum\nolimits_{j,k=0}^{N}\hat{\uB}^2_{\text{Leg},ijk}\right)^{i=m}}{\sum\nolimits_{i,j,k=0}^{N}\hat{\uB}^2_{\text{Leg},ijk}}\right].
    \label{eq:modal_contribution}
\end{equation}
Subsequently, the decay rate $\sigma^i$ in the $i^\text{th}$ direction is obtained with a least-squares fit of 
the contributions $w_m^{\xi_i}$ to an exponential function
\begin{equation}
    w_m^{\zeta_i}=c e^{-\sigma^i m} \quad \text{with} \quad c\in\mathbb{R}. 
    \label{eq:modal_fit}
\end{equation}
The final indicator value is obtained as the minimum over all spatial dimensions 
\begin{equation}
    \mathcal{I} = \underset{\Omega}{\text{min}}(\lvert\sigma_1\rvert,\lvert\sigma_2\rvert,\lvert\sigma_3\rvert).
    \label{eq:modal_indicator}
\end{equation}

The indicator function $\mathcal{I}$ serves two purposes: it controls FV shock capturing and determines the local ansatz degree in 
the p-adaptive DG operator. Switching between ansatz degrees and shock capturing relies on evaluating and comparing $\mathcal{I}$ against 
empirical thresholds $\mathcal{T}$, as detailed in~\cite{Mossier2022,Mossier2023}. To detect and distinguish solution features that 
require either shock capturing or p-refinement, two control mechanisms are available: first, the selection of variables on which $\mathcal{I}$ 
is evaluated -- pressure $p$ for shocks, density $\rho$ for shear layers, and mass fraction $\YB$ for material interfaces -- and the tuning of   
the thresholds $\mathcal{T}$, which are separately defined for shock capturing and p-refinement. 
Since shocks and sharp gradients in the species concentration can trigger spurious oscillations, 
shock capturing relies on $\mathcal{I}\left(p\right)$ and $\mathcal{I}\left(\YB\right)$. On the other hand, p-adaptation
is controlled by $\mathcal{I}\left(\rho\right)$ to track contact discontinuities at shear layers. 
Hence, the indicators $\mathcal{I}_{\text{p-adapt}}$ for p-adaptation and $\mathcal{I}_{\text{h-refine}}$ for h-refined sub-cell limiting are implemented as follows:
\begin{equation}
	\mathcal{I}_{\text{p-adapt}}:=\mathcal{I}\left(\rho\right), \quad \mathcal{I}_{\text{h-refine}}\left(p,\YB\right):=\text{min}(\mathcal{I}\left(p\right),\mathcal{I}\left(\YB\right)).
\end{equation} 

\subsection{Dealiasing and Limiting}
\label{subsec:dealiasing_filtering}

The numerical quadrature of the non-linear flux function, using collocation, introduces
integration errors, which may lead to aliasing and thus degrade stability in the presence of under-resolved turbulence. To overcome 
this issue, we apply the non-linearly stable split-flux form of Gassner et al.~\cite{Gassner2016}, which is derived from the strong 
form of the DGSEM on LGL nodes. The split-form is constructed such that it fulfills the summation-by-parts (SBP) property, a discrete
analogue to integration by parts. Using the SBP property and replacing the fluxes by suitable two-point fluxes,
the split-form \DG allows to preserve properties like kinetic energy at a discrete level, independently of the quadrature error.  
In the present paper, kinetic energy preserving two-point fluxes of Pirozzoli~\cite{Pirozzoli2011} are employed. The split-operator
can be written in an equivalent form to the semi-discrete operators \eqref{eq:semidiscrete}, albeit with
the fluxes of the Navier--Stokes equations replaced by two-point fluxes~\cite{flexi}. Therefore, the coupling strategy at element interfaces 
is unaffected by the choice between the standard DGSEM or split-form DGSEM.  

While the split-form and the \FV shock-capturing greatly enhance robustness and shock-localization, a formal guarantee of
stability is impossible in the face of empirically tuned indicator thresholds 
$\mathcal{T}$ and even a second-order \FV reconstruction in multiple space-dimensions. 
Thus, we additionally apply the positivity preserving (PP) limiter of Zhang and Shu~\cite{Shu2010} to ensure 
positive density and pressure during computations. The relative amount of elements affected by the
PP-limiter will be addressed in the numerical application section, to demonstrate that this additional stabilization is indeed only 
required for a negligible number of elements. 

\subsection{Dynamic Load Balancing}
\label{subsec:dynamicLoadBalancing}

Dynamic hp-adaptation at runtime introduces variable computational costs per element and thus workload imbalances among 
processors in parallel execution. 
For an efficient implementation of the adaptive discretization on massively parallel high-performance computers, 
an efficient dynamic load balancing (DLB) scheme is essential. DLB ensures an even distribution of computational workloads 
throughout the computation and relies on three central building blocks:
\begin{enumerate}
	\item \label{item:workload} An estimation of the current workload per processor unit $\mathcal{P}$. 
	\item \label{item:partitioning} A partitioning algorithm that divides the domain $\Omega$ into subdomains $\Omega_{\mathcal{P}}$,
	ensuring a balanced workload distribution across the processors $\mathcal{P}_n$ for $n\in[1,N_\text{procs}]$.
	\item \label{item:heuristic} A heuristic for selecting a set of time steps $\sigma=\{t_0,t_1,...,t_{\text{end}}\}$,
	referred to as a scenario, at which dynamic load balancing is to be performed.
\end{enumerate}

In the present work, we utilize the DLB implementation proposed in~\cite{Mossier2023,MossierPHD}. 
The method estimates the workload per element for each discretization through an initial calibration run at the start of the computation.
During this calibration, the spatial operator is evaluated once for a purely finite volume 
discretization $\smash{\bm{\mathcal{O}}^{FV}}$ and for discontinuous Galerkin discretizations $\smash{\bm{\mathcal{O}}^{DG(N)}}$ with polynomial degrees 
$N=N_{\text{min}},...,N_{\text{max}}$, yielding the element weights 
\begin{equation}
	w(\bm{\mathcal{O}}^{FV}),\, w(\bm{\mathcal{O}}^{DG(N_{\text{min}})}),\, ...,\, w(\bm{\mathcal{O}}^{DG(N_{\text{max}})}).
\end{equation}
The workload per processor $B_{n}$ is then obtained as the sum of the element weights $\smash{B_{n}=\sum^{N_{\text{elems}}}_{i=1}w_i}$ over the sub-partitions,
assigned to the respective processor $\mathcal{P}_n$.

Partitioning of the domain $\Omega$ into subdomains $\Omega_{\mathcal{P}}$ is achieved by solving a chains-on-chains partitioning problem~\cite{Pinar2004} 
along a space filling curve.
The quality of the partitioning can be described by the imbalance $\mathcal{I}$, which is defined as 
\begin{equation}
	\mathcal{I}=\underset{n\in[1,N_\text{procs}]}{\max} \left( \frac{B_{n}}{B^*}-1 \right) \quad \mathcal{I}\in(0,\infty]
\end{equation}
with the target weight per partition $B^*$. Thereby, a value of $\mathcal{I}=0$ corresponds to an ideal partition.

Finally, a scenario $\sigma$ is chosen based on a simple heuristic where DLB is performed at most every $\mathcal{S}_{\text{DLB}}$ time steps, 
when the imbalance exceeds a threshold of $\mathcal{I}>\mathcal{T}_{\text{DLB}}$. The performance of the described DLB strategy is discussed
in Section \ref{sec:Application} with a study of the imbalances and a scaling test. 

\section{Numerical Validation}
\label{sec:Validation}
Within this section, key properties of the proposed numerical scheme like free-stream preservation and the experimental order
of convergence are analyzed. Further, the method is applied to the well-know compressible Taylor-Green Vortex benchmark to investigate its capability 
in handling under-resolved compressible turbulence. Finally, we study a multi-component shock-triplepoint interaction on a 
non-conforming grid. 

\subsection{Free-Stream Preservation}
\label{subsec:freestream}
\begin{figure}[t]
	\tikzsetnextfilename{freestream_domain}
	\input{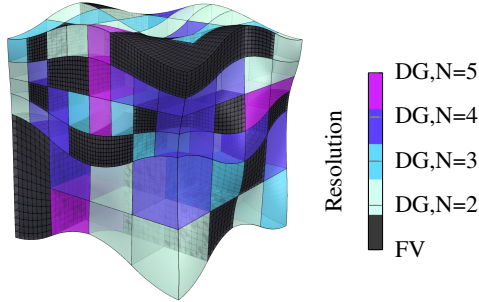}
	\caption{Distorted cubic domain of the free-stream preservation test. The upper half uses
	a higher resolution, leading to a non-conforming mesh. A random distribution of \DG and \FV elements 
	is used with an ansatz degree $N=[2,5]$ and an \FV sub-cell resolution of $N_{\text{FV}}=8$ per direction.}
	\label{fig:freestream}
\end{figure}

This section validates the proposed hp-adaptive discretization by demonstrating its ability 
to preserve a uniform free-stream solution on a curved, non-conforming grid with varying local element discretizations.
An initial free-stream solution $\uB=\left(1,1,1,1,1\right)^T$ is imposed on a computational domain $\Omega$, which is defined 
as a distorted periodic cube with the mapping:
\begin{equation}
	\xB \mapsto \xB+0.1 \cdot \text{sin}(x_1) \cdot \text{sin}(x_2) \cdot \text{sin}(x_3), \quad \text{for}\; \xB\in[-1,1]^3.  
	\label{eq: freestream_domain}
\end{equation}
The lower half of the cube features twice the resolution than the upper half, resulting in a non-conforming grid.
A polynomial ansatz of degree $N_{geo}=2$ is chosen to approximate the physical geometry and the solution is discretized with 
a variable ansatz degree $N\in\left[2,5\right]$ and \FV elements with a sub-cell resolution of $N_{\text{FV}}=8$.
To ensure the most rigorous testing, we distribute the ansatz degree and \FV elements randomly in space and time. The resulting 
setup is visualized in Figure \ref{fig:freestream}. The simulation is conducted for $t\in[0,0.5]$ which amounts to $180$ time steps.
A subsequent analysis of the $\mathbb{L}_2$ and $\mathbb{L}_{\infty}$ error norms, listed in Table \ref{table:freestream}, indicates 
errors are within the order of machine precision. Thus, we can conclude that the proposed dynamic adaptive scheme achieves free-stream 
preservation on non-conforming grids.

\begin{table}[!h]
	\centering
	\begin{tabular}{c|c|c|c|c|c}
		& $u_1$		& $u_2$ 	& $u_3$ 	& $u_4$ 	& $u_5$ \\  
		\hline
        \rule{0pt}{2ex}   
		$\mathbb{L}_2$  		 & $1.42e^{-14}$	& $1.13e^{-14}$ 	& $1.15e^{-14}$ 	& $1.16e^{-14}$ 	& $4.36e^{-14}$ \\ 
		$\mathbb{L}_{\infty}$    & $3.91e^{-14}$	& $2.88e^{-14}$ 	& $2.96e^{-14}$ 	& $3.03e^{-14}$ 	& $9.81e^{-14}$ \\ 
		\hline                                                   
	\end{tabular}
	\caption{$L_2$ and $L_{\infty}$ free-stream error with a hybrid \DGFV discretization after 180 time steps.}
	\label{table:freestream}	
\end{table}

\subsection{Experimental Order of Convergence}
\label{subsec:convergence}
To establish the proposed method on non-conforming grids, we show an experimental analysis of the error convergence. 
The convergence study considers a smooth density wave, which is transported diagonally along the axis $\left[1,1,1\right]^T$ 
within a distorted cubic domain analogue to that in Section \ref{subsec:freestream}. 

\begin{figure}[t]
	\tikzsetnextfilename{domain}
	\input{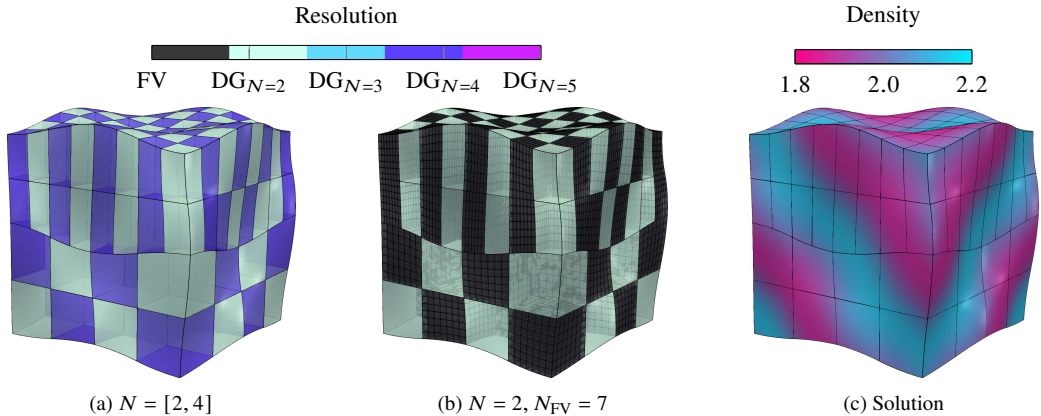}
	\caption{Distorted cubic domain of the convergence test. Figures \ref{subfig:p_conv} and Figure \ref{subfig:h_conv} depict
	the checkerboard distribution of ansatz degrees and \FV elements respectively. Figure \ref{subfig:conv_sol} shows the density wave, which is
	propagated through the domain.}
	\label{fig:convtest_domain}
\end{figure}

A non-conforming discretization 
is chosen by prescribing twice the resolution in the $x_1$ and $x_2$ directions for the lower half of the domain, $x_3<0$. Figure 
\ref{fig:convtest_domain} illustrates the discretized domain $\Omega$ and the solution at $t=0$.

\begin{figure}[b!]
	\centering
	\tikzsetnextfilename{convtest}
	\begin{tikzpicture}[font=\small,scale=0.90]

    \begin{groupplot}[
          group style={
            group size=2 by 1,
            horizontal sep={0.09\linewidth},
            vertical sep={0.06\linewidth}
          },
          legend style={
            at={(-0.2,-0.27)},  %
            anchor=north,     %
            legend columns=4, %
            /tikz/every even column/.append style={column sep=0.7cm} %
          },
          ylabel near ticks,xlabel near ticks,
          width=\linewidth
      ]
    
      \def\figwidth{0.49\linewidth} %
      \def\figheight{0.47\linewidth} %
      \def\ycol{1}    %
    
      \nextgroupplot[%
        width=\figwidth,
        height=\figheight,
        title={Standard DG -- LG nodes},
        xlabel={$\Delta x$},
        ylabel={$L_2$-Error $\left(\rho\right)$},
        ymin=2.e-9,ymax=1.e-1,
        ymode=log, xmode=log,
        grid=both, %
        major grid style={line width=0.4pt, gray!50}, %
        minor grid style={line width=0.2pt, gray!30},  %
        colormap name=cool,          %
        cycle list={{cyan!80!black}, {blue!80!white}, {purple!70!blue}, {teal!80!black}}
      ]
      \addplot+[thick,mark=square*] table[x expr = 2/((\thisrowno{0})), y expr=\thisrowno{1}, col sep=comma,skip first n=1]{./pictures/convergence/non_conforming_curved/DG_N_24/convtest_p_dg_N24_hconvfile_N2.csv};
      \addplot+[thick,mark=square*] table[x expr = 2/((\thisrowno{0})), y expr=\thisrowno{1}, col sep=comma,skip first n=1]{./pictures/convergence/non_conforming_curved/DG_N_35/convtest_p_dg_N35_hconvfile_N3.csv};
      \addplot+[thick,mark=square*] table[x expr = 2/((\thisrowno{0})), y expr=\thisrowno{1}, col sep=comma,skip first n=1]{./pictures/convergence/non_conforming_curved/DG_N_46/convtest_p_dg_N46_hconvfile_N4.csv};
      \addplot+[thick,mark=square*] table[x expr = 2/((\thisrowno{0})), y expr=\thisrowno{1}, col sep=comma,skip first n=1]{./pictures/convergence/non_conforming_curved/DG_N_23_FV_6/convtest_dg_fv_N23_fv_6_hconvfile_N2.csv};

      \node at (0.07, 16e-5) (d) {};
      \node at (0.07,64e-5) (e) {};
      \node at (0.14,64e-5) (f) {};
      \draw[black!80] (d.center) -- (e.center) -- (f.center) -- (d.center);
      \node[anchor=south] at ($(e)!0.5!(f)$) {\footnotesize$1$};
      \node[anchor=east]  at ($(d)!0.5!(e)$) {\footnotesize$2$};

      \node at (0.11, 8e-6) (a) {};
      \node at (0.11,64e-6) (b) {};
      \node at (0.055, 8e-6) (c) {};
      \draw[black!80] (a.center) -- (b.center) -- (c.center) -- (a.center);
      \node[anchor=north] at ($(a)!0.5!(c)$) {\footnotesize$1$};
      \node[anchor=west]  at ($(a)!0.5!(b)$) {\footnotesize$3$};
    
      \node at (0.2,  4e-8) (a) {};
      \node at (0.2,128e-8) (b) {};
      \node at (0.1,  4e-8) (c) {};
      \draw[black!80] (a.center) -- (b.center) -- (c.center) -- (a.center);
      \node[anchor=north] at ($(a)!0.5!(c)$) {\footnotesize$1$};
      \node[anchor=west]  at ($(a)!0.5!(b)$) {\footnotesize$5$};
    
      \nextgroupplot[%
      width=\figwidth,
      height=\figheight,
      title={Split-Flux DG -- LGL nodes},
      xlabel={$\Delta x$},
      ymin=2.e-9,ymax=1.e-1,
      ymode=log, xmode=log,
      grid=both, %
      major grid style={line width=0.4pt, gray!50}, %
      minor grid style={line width=0.2pt, gray!30},  %
      colormap name=cool,          %
      cycle list={{cyan!80!black}, {blue!80!white}, {purple!70!blue}, {teal!80!black}}
    ]
    \addplot+[thick,mark=square*] table[x expr = 2/((\thisrowno{0})), y expr=\thisrowno{1}, col sep=comma,skip first n=1]{./pictures/convergence/non_conforming_curved_split/DG_N_24/convtest_p_dg_N24_hconvfile_N2.csv};
    \addplot+[thick,mark=square*] table[x expr = 2/((\thisrowno{0})), y expr=\thisrowno{1}, col sep=comma,skip first n=1]{./pictures/convergence/non_conforming_curved_split/DG_N_35/convtest_p_dg_N35_hconvfile_N3.csv};
    \addplot+[thick,mark=square*] table[x expr = 2/((\thisrowno{0})), y expr=\thisrowno{1}, col sep=comma,skip first n=1]{./pictures/convergence/non_conforming_curved_split/DG_N_46/convtest_p_dg_N46_hconvfile_N4.csv};
    \addplot+[thick,mark=square*] table[x expr = 2/((\thisrowno{0})), y expr=\thisrowno{1}, col sep=comma,skip first n=1]{./pictures/convergence/non_conforming_curved_split/DG_N_23_FV_6/convtest_dg_fv_N23_fv_6_hconvfile_N2.csv};
  
    \addlegendentry{$N=[2,4]$}
    \addlegendentry{$N=[3,5]$}
    \addlegendentry{$N=[4,6]$}
    \addlegendentry{$N=2,N_{\text{FV}}=7$}

    \node at (0.07, 3.5*16e-5) (d) {};
    \node at (0.07,3.5*64e-5) (e) {};
    \node at (0.14,3.5*64e-5) (f) {};
    \draw[black!80] (d.center) -- (e.center) -- (f.center) -- (d.center);
    \node[anchor=south] at ($(e)!0.5!(f)$) {\footnotesize$1$};
    \node[anchor=east]  at ($(d)!0.5!(e)$) {\footnotesize$2$};

    \node at (0.11, 2*8e-6) (a) {};
    \node at (0.11,2*64e-6) (b) {};
    \node at (0.055, 2*8e-6) (c) {};
    \draw[black!80] (a.center) -- (b.center) -- (c.center) -- (a.center);
    \node[anchor=north] at ($(a)!0.5!(c)$) {\footnotesize$1$};
    \node[anchor=west]  at ($(a)!0.5!(b)$) {\footnotesize$3$};
  
    \node at (0.2,  1.5*4e-8) (a) {};
    \node at (0.2,1.5*128e-8) (b) {};
    \node at (0.1,  1.5*4e-8) (c) {};
    \draw[black!80] (a.center) -- (b.center) -- (c.center) -- (a.center);
    \node[anchor=north] at ($(a)!0.5!(c)$) {\footnotesize$1$};
    \node[anchor=west]  at ($(a)!0.5!(b)$) {\footnotesize$5$};

    \end{groupplot}
    \end{tikzpicture}%
    
	\caption{Convergence of the hp-adaptive operator with the split-flux \DG scheme on Legendre--Gauss nodes (\textit{left}) and the standard \DG scheme 
	on Legendre--Gauss--Lobatto nodes (\textit{right}).}
	\label{fig:convtest_result}
\end{figure}
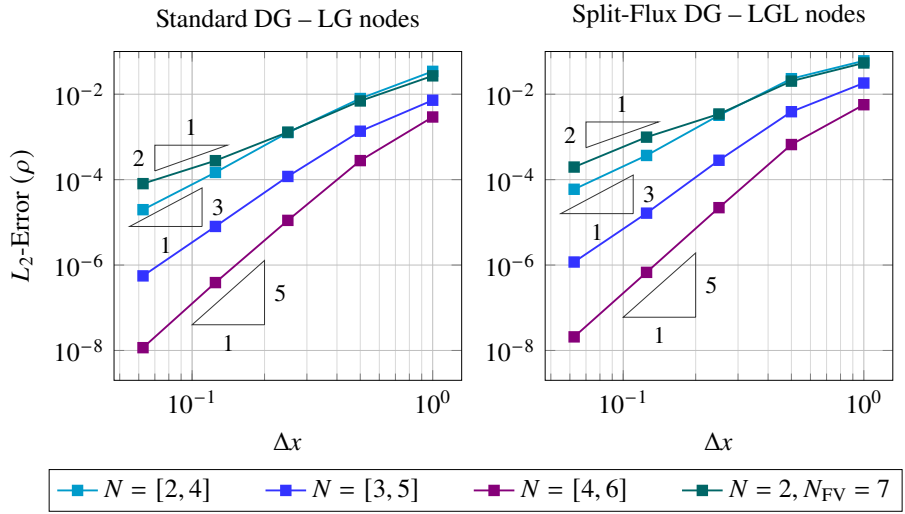

To obtain a representative study of the hp-adaptive operator, different combinations of element local discretizations are chosen: 
first, the p-adaptive \DG operator is applied for three checkerboard-like distributions of the ansatz degree $N=[2,4]$, $N=[3,5]$ and $N=[4,6]$. 
Then, a hybrid \DGFV discretization is chosen with \FV elements positioned in a checkerboard pattern and a resolution of $N=2$ and $N_{\text{FV}}=7$.
To avoid interference of the temporal discretization error, the CFL number is reduced to $0.5$. 
The convergence study
is performed for the standard \DG scheme on Legendre--Gauss nodes and repeated for the split-flux \DG scheme on Legendre--Gauss--Lobatto nodes. 
In Figure \ref{fig:convtest_result}, the resulting $\mathbb{L}_{2}$ error norms in the density, $\rho$, are plotted as a function of the grid size $\Delta x$.
A comparison against the expected order of convergence $p=N+1$ of the truncation error $\smash{e\propto k(\Delta x)^P}$ for a smooth polynomial discretization
of degree $N$ indicates the desired convergence behavior has been achieved. Consequently, the convergence order matches the lowest present discretization 
in the setup for all considered combinations. 

\subsection{Compressible Taylor-Green Vortex}
\label{subsec:tgv}
A well-know challenge for high-order methods is to achieve both accuracy and robustness in the presence of under-resolved turbulence and shocks. 
In the context of the present hp-adaptive scheme, a key requirement is the reliable distinction between turbulent structures, which are to be treated by
p-refinement, and shocks, where the h-refined \FV sub-cell operator is needed. 
With the supersonic Taylor-Green vortex problem, a benchmark has been established~\cite{Lusher2021,Chapelier2024} to assess the ability of numerical methods 
to deal with progressively decreasing turbulent scales in the presence of shocks. The supersonic TGV is defined in a periodic, cubic domain 
$\Omega=[-\pi,\pi]^3$ with the  isothermal initial conditions 
\begin{subequations}
	\begin{align}
	v_1\left(\xB,0\right)&=\text{sin}\left(x_1\right)\text{cos}\left(x_2\right)\text{cos}\left(x_3\right), \\
	v_2\left(\xB,0\right)&=-\text{cos}\left(x_1\right)\text{sin}\left(x_2\right)\text{cos}\left(x_3\right), \\
	v_3\left(\xB,0\right)&=0.0, \\
	p\left(\xB,0\right)  &=p_0+\frac{\rho_0}{16}\left(text{cos}(2x_1)+\text{cos}(2x_2)\right)\left(2+\text{cos}(2x_3)\right),
	\end{align}
\end{subequations}
where the temperature $T_0$ is set to the reference temperature of the Sutherland law $T_{\text{ref}}$ and a ratio of heat capacities 
$\gamma=1.4$. To close the initial conditions, the Mach number is set to $\smash{\text{Ma}_0=\sqrt{\frac{\rho_0}{\gamma p_0}}\coloneqq 1.25}$ and the Reynolds number chosen as 
$\smash{\text{Re}_0 = \frac{\rho_0}{\mu_0}\coloneqq 1600}$. A non-constant dynamic viscosity is applied, following the Sutherland law, as 
explained in~\cite{Chapelier2024}.

The TGV problem is advanced until a final time $t=20$ and the physical domain $\Omega$ is discretized by $32^3$ elements. 
The hp-adaptive discretization applies a variable ansatz degree $N\in[2,4]$
and an \FV sub-cell resolution of $N_{\text{FV}}=9$. To distinguish between shocks and contact discontinuities, the indicator for p-refinement is evaluated
on the pressure, while \FV shock-capturing is controlled by the indicator, operating on the density. The computation employs the standard
DGSEM on LG nodes and does not require the PP-limiter from Section \ref{subsec:dealiasing_filtering}. 

\begin{figure}[t!]
	\includegraphics[width=1.0\textwidth]{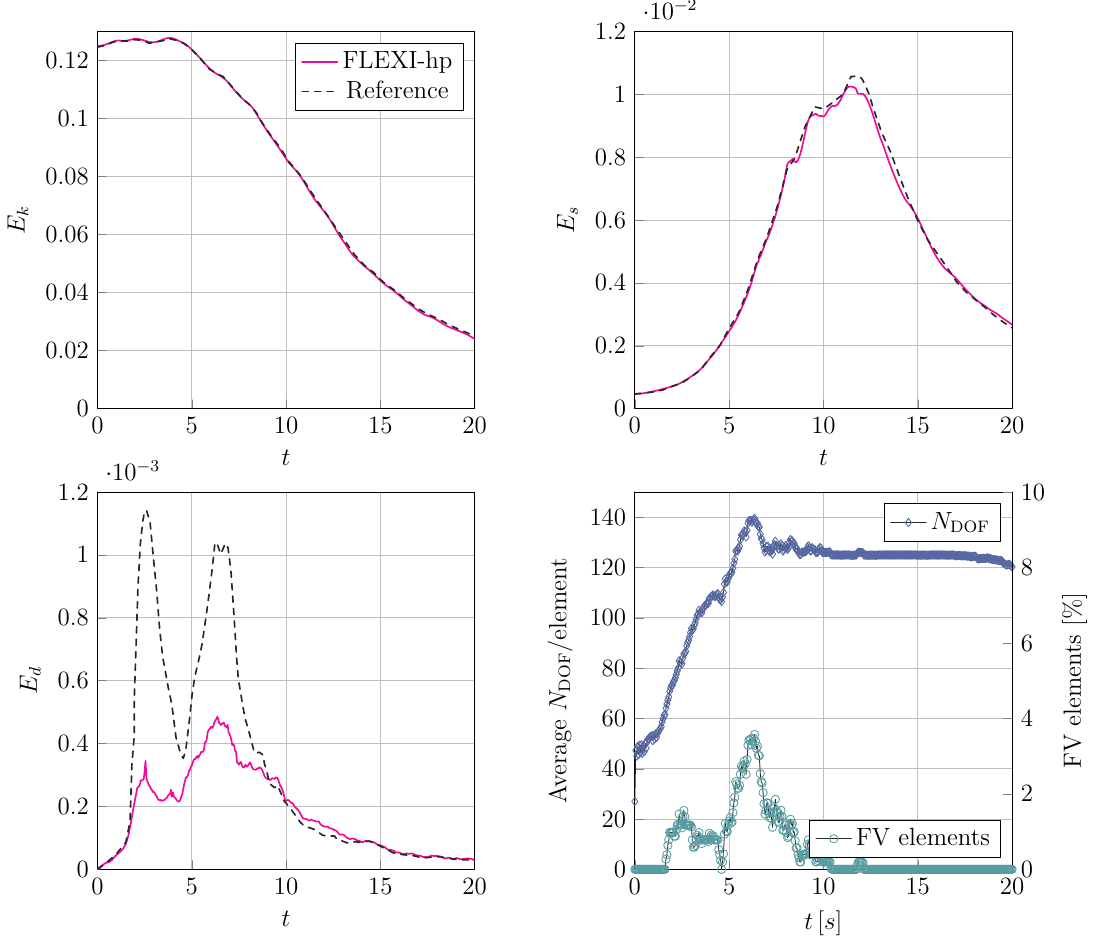}
	\caption{Results of the TGV flow at $\text{Re}_0=1600$ and $\text{Ma}_0=1.25$ for $t=[0,20]$. The plots evaluate the temporal evolution of 
	the kinetic energy $E_k$ and the solenoidal and dilatational components of the kinetic energy dissipation $E_s$ and $E_d$.
	In addition, a statistic of the relative amount of \FV elements and the average number of DOFs per element $N_{\text{DOF}}$ is provided.}
	\label{fig:TGV}
\end{figure}

Results obtained with the present scheme are compared against the reference solution from~\cite{Chapelier2024}, which was computed with a $6^{th}$-order
Targeted Essentially Non-Oscillatory (TENO) scheme on a very fine mesh of $2048^3$ DOFs. Following~\cite{Lusher2021,Chapelier2024}, we evaluate the solution based on three quantities: the kinetic energy 
$E_k$ describing the energy carried by the turbulent motion, the solenoidal part of the kinetic energy dissipation $E_s$, which is related 
to the vortical motion in the flow, and the dilatational component of the kinetic energy dissipation $E_d$, which correlates the onset of shocks in the flow.
In Figure \ref{fig:TGV}, we can observe a near perfect agreement of $E_k$ and a close agreement in $E_s$. Though $E_d$ deviates significantly from the 
reference solution, it is in good agreement with computations of similar resolution shown in~\cite{Chapelier2024}.
To assess the performance of the indicator scheme, Figure \ref{fig:TGV} also provides a statistic of the amount of \FV sub-cells and
the average number of DOFs per element. While the amount of \FV sub-cells never exceeds $4\%$, we want to emphasize the near perfect correlation
between the amount of \FV cells and $E_d$. Both peaks in $E_d$ at $2.5s$ and $7s$ are perfectly mirrored in time trace of the amount of \FV cells. 
Similarly, the average number of DOFs
per element increases with $E_s$ until it caps at $t=12$ since the maximum allowed ansatz degree prohibits further refinement. 
In conclusion, we state that the hybrid adaptive scheme proves to be well-suited to deal with both under-resolved turbulence and shocks.  

\subsection{Shock-Triplepoint Interaction}
\label{subsec:sti}
Finally, the multi-component implementation is validated with a two-component shock-triplepoint interaction problem~\cite{Loubere2010,Breil2013}. It is defined as a three-state 
Riemann problem in the two-dimensional domain $\Omega=[0,7]\times[-1.5,1.5]$ with the initial conditions:
\begin{equation*}
	\uB(\xB,t=0) = 
	\begin{cases}
		\rho = 1.0   \;, & p = 1.0 , \gamma = 1.4 \quad \text{if} \;\xB \in \Omega_1=[0,1]\times[-1.5,1.5] \\
		\rho = 1.0   \;, & p = 0.1 , \gamma = 1.4 \quad \text{if} \;\xB \in \Omega_2=[1,7]\times[-1.5,0]   \\
		\rho = 0.125 \;, & p = 0.1 , \gamma = 1.5 \quad \text{if} \;\xB \in \Omega_3=[1,7]\times[0,1.5].
	\end{cases}
\end{equation*}
Both viscosity and heat conduction are neglected, while a constant species diffusion coefficient $D_k=1\cdot10^{-6}$
is assumed for the species $k\in[1,2]$.

The initial pressure jump results in a shock that propagates with different speeds in the subdomains $\Omega_2$ and $\Omega_3$,
which causes the formation of a distinctive shear layer at the material interface between the fluid components. 
\begin{figure}[t]
	\centering
	\includegraphics[width=0.7\textwidth]{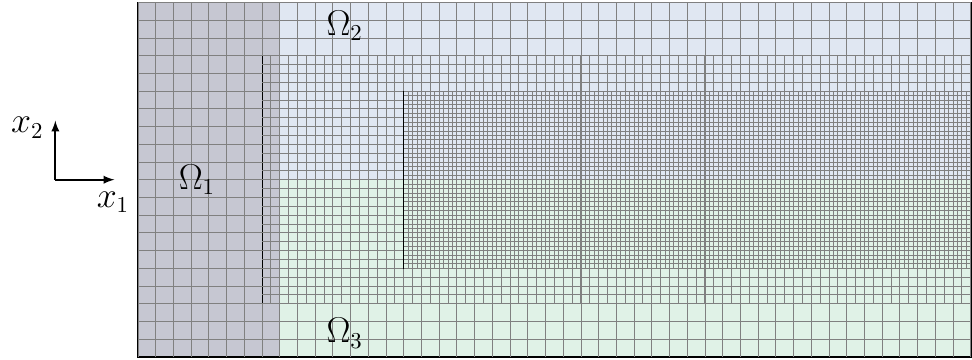}
	\caption{Non-conforming mesh for the shock-triplepoint interaction with two refinement levels.}
	\label{fig:setup_sti}
\end{figure}
The problem is simulated for $t\in[0,6]$ within the domain $\Omega$ and discretized by a non-conforming grid 
with a base resolution of $280\times120$ elements and two refinement levels up to an effective resolution of $1120\times480$.

\begin{figure}[htb!]
    \centering
    \tikzsetnextfilename{results}
    \input{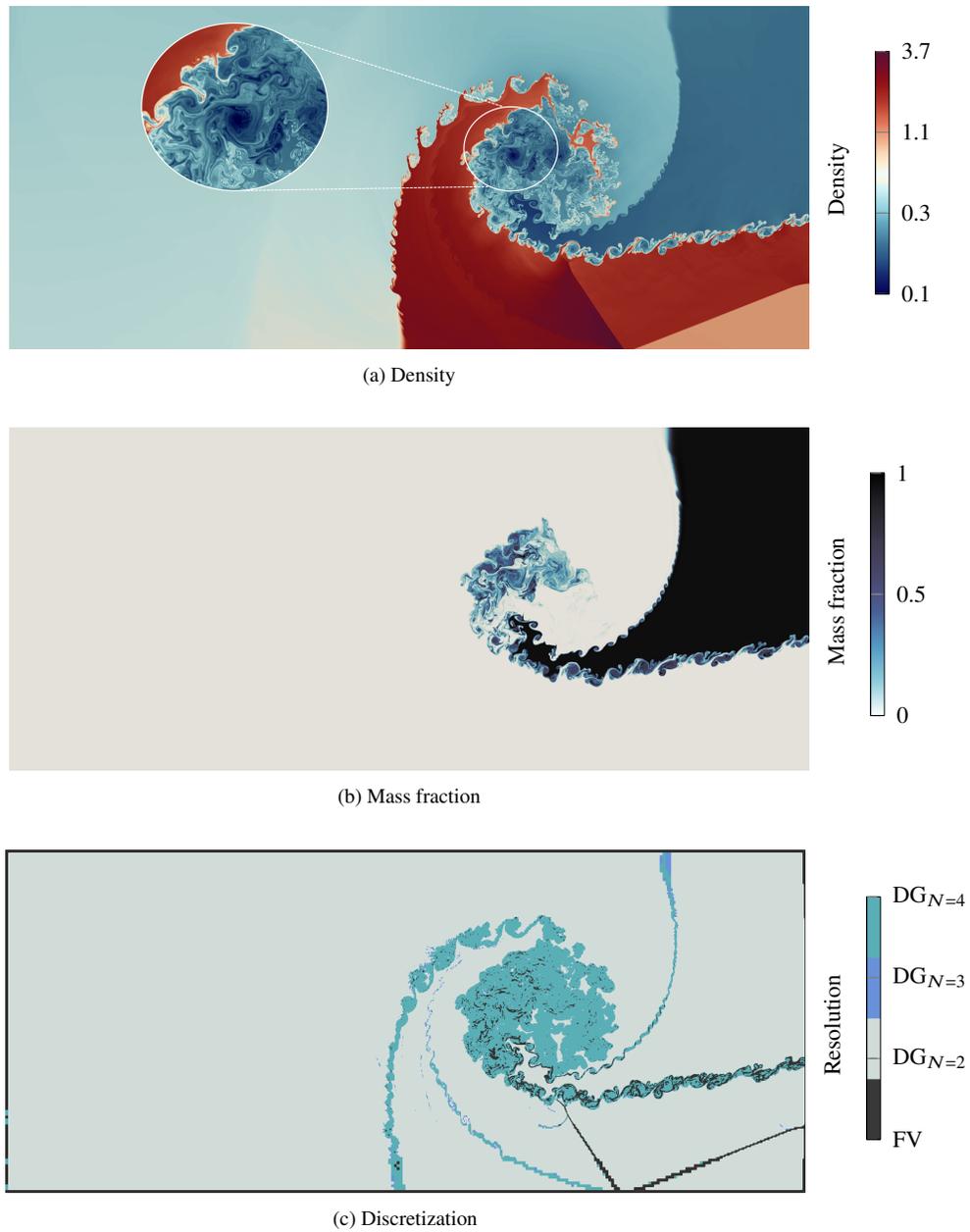}
    \caption{Multi-component simulation of a two-phase shock-triplepoint interaction problem. The subplots provide the density field (a) and mass 
	fraction (b) and indicate the local resolution (c) at the time instance $t=5.0$. }
    \label{fig:shock_triple_point}
\end{figure}

In total, the discretization uses roughly $0.25$ million \DG elements. Inflow conditions are imposed on the left boundary, while 
all remaining boundaries apply non-reflecting outflow conditions.
Figure \ref{fig:setup_sti} provides an illustration of the geometric setup. 
The problem is computed with a p-adaptive \DG operator with $N=[2,4]$ and an \FV sub-cell resolution of $N_{\text{FV}}=9$.
As in the TGV problem, p-refinement relies on the indicator evaluated on the density and shock-capturing is triggered
by the indicator operating on the pressure. 
However, since material interfaces are prone to produce oscillations in high-order methods, the indicator is additionally
evaluated on the mass fraction to stabilize sharp gradients in the concentration with the \FV operator. 
Due to the lack of viscosity, the formation of intricate vortical structures in the shear layers 
requires the added non-linear stability of the split-form DGSEM, as discussed in Section \ref{subsec:dealiasing_filtering}.   

Figure \ref{fig:shock_triple_point} illustrates the flow field (\ref{fig:shock_triple_dens}), 
mass fraction (\ref{fig:shock_triple_frac}) and the element local discretization (\ref{fig:shock_triple_hp})
at the time instance $t=5$. Both the density field and mass fraction show the expected results, with 
rich vortical structures emanating from the entropy shear layers and the roll-up of the material interface. 
Figure \ref{fig:shock_triple_hp} demonstrates an excellent performance of the indicators with precise
shock detection and p-refinement applied at all primary and even secondary shear layers. Moreover, the
indicator evaluation on the mass fraction successfully detects
areas with strong concentration gradients and increases stability there by placing \FV sub-cells.  
Here, the advantage of the increased \FV sub-cell resolution $N_\text{FV}>N_{\text{max}}$ is demonstrated:
the h-refined sub-cell grid compensates for the reduced order with an increased number of DOFs per element. Consequently,
there is no extensive numerical dissipation apparent in the shear layer at the material interface,
where FV stabilization of strong concentration gradients is required. 
Finally, it can be remarked, that the mortar interfaces do not produce any visible artifacts in the solution, 
though their influence is apparent in the top and bottom of Figure \ref{fig:shock_triple_dens} where contact lines are somewhat smeared out
in the less refined regions. 

\section{Application - Under-Expanded $\text{H}_2$ Injection Jet}
\label{sec:Application}
As a challenging application for the proposed adaptive scheme, a scale-resolving simulation of a supersonic hydrogen jet 
is performed. In the face of the climate crisis and the depletion of fossil fuel reserves, hydrogen has emerged as a 
promising alternative for internal combustion engines.
The high-speed injection of hydrogen into an air atmosphere plays a critical role in direct hydrogen injection engines.
However, unlike conventional liquid fuels, hydrogen's compressibility and low density necessitate high-pressure injection  
to achieve sufficient mass flow rates. This leads to a choked nozzle flow characterized by shocks and a supersonic, 
under-expanded jet featuring distinct shock diamonds.
The formation and breakup of this jet governs the mixing of hydrogen fuel with air, 
making a detailed understanding of this process essential for optimizing combustion efficiency.
Simulating such flows presents significant challenges due to the nonlinear interactions of shock waves, 
turbulence, and acoustics, which span a wide range of spatial and temporal scales.

\subsection{Numerical Setup}
\begin{figure}[b!]
	\begin{subfigure}[b!]{0.49\textwidth}
		\begin{tikzpicture}
			\node[anchor=south west, inner sep=0] (image) at (0,0) {\includegraphics[width=5.5cm]{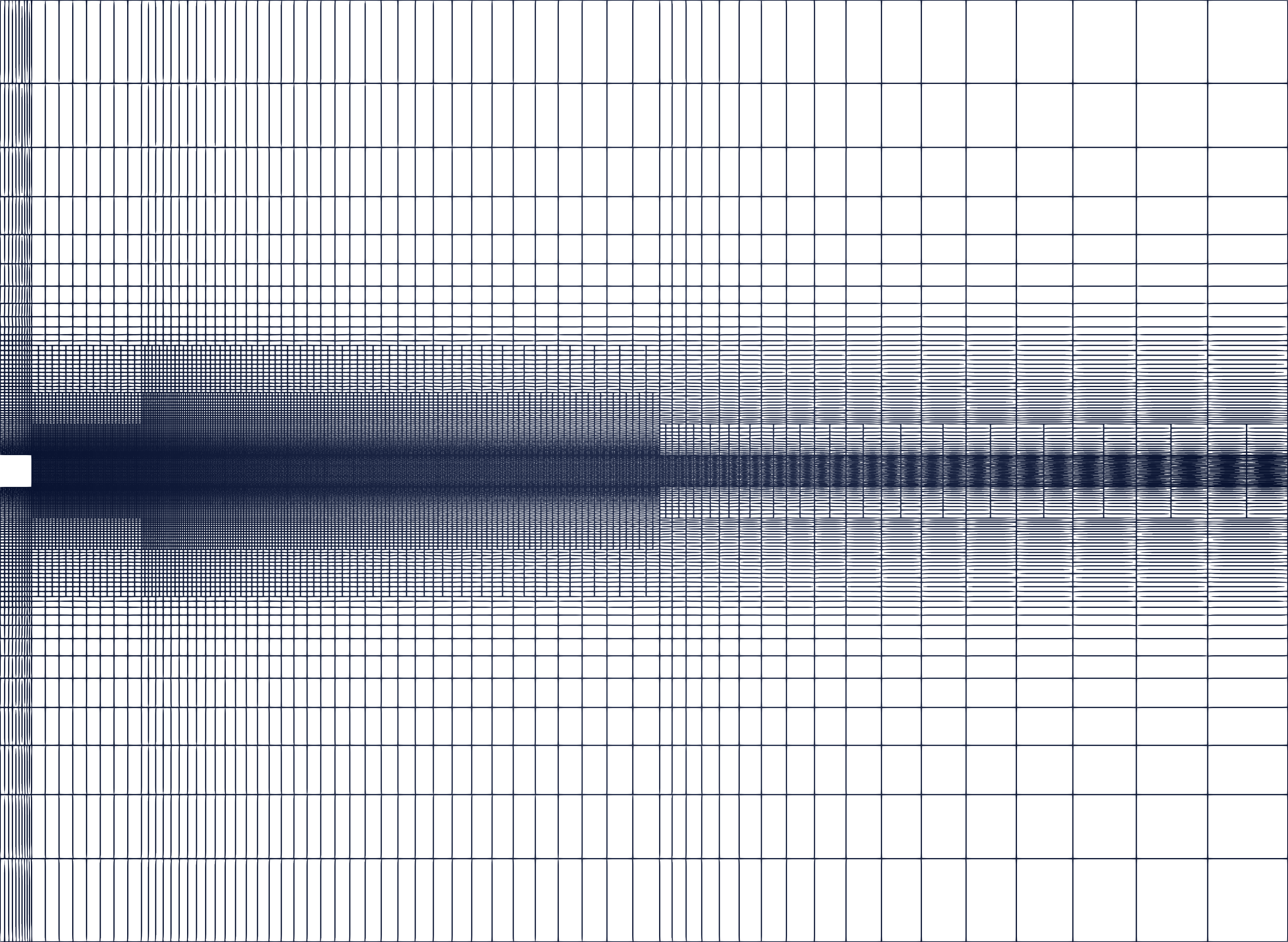}};
			\begin{scope}[x={(image.south east)},y={(image.north west)}]
				\definecolor{pink1}{RGB}{240,50,200}
				\draw[<->, black, thick] (1-0.9756, -0.07) -- (1, -0.07) node[midway, below] {\scriptsize $40 D_0$};
				\draw[<->, black, thick] (-0.07, -0.07) -- (1-0.9756, -0.07) node[midway, below] {\scriptsize $D_0$};
		
				\draw[-, gray, dashed] (-0.07, -0.065)  -- (0.0, -0.005);
				\draw[-, gray, dashed] (1-0.9756, -0.065) -- (1-0.9756,-0.005);
		
				\draw[<->, black, thick] (1.05, 0.0) -- (1.05, 1.0) node[midway, right] {\scriptsize $30 D_0$};
		
				\draw[<->, black, thick] (-0.07, 0.5+0.1) -- (-0.07, 0.5-0.1) node[midway, left] {\scriptsize $D_0$};
				\draw[-, gray, dashed] (-0.060, 0.5+0.1)  -- (-0.005, 0.5+0.0167);
				\draw[-, gray, dashed]  (-0.060, 0.5-0.1) -- (-0.005, 0.5-0.0167);
				\draw[dashed, pink1,thick] (0.0244, -0.03) -- (0.0244, 1.03);
			\end{scope}
		\end{tikzpicture}
		\caption{2D Slice}
		\label{subfig:mesh_visu_2D}
	\end{subfigure}
	\hfill
	\begin{subfigure}[b!]{0.43\textwidth}
		\includegraphics[width=1.0\textwidth]{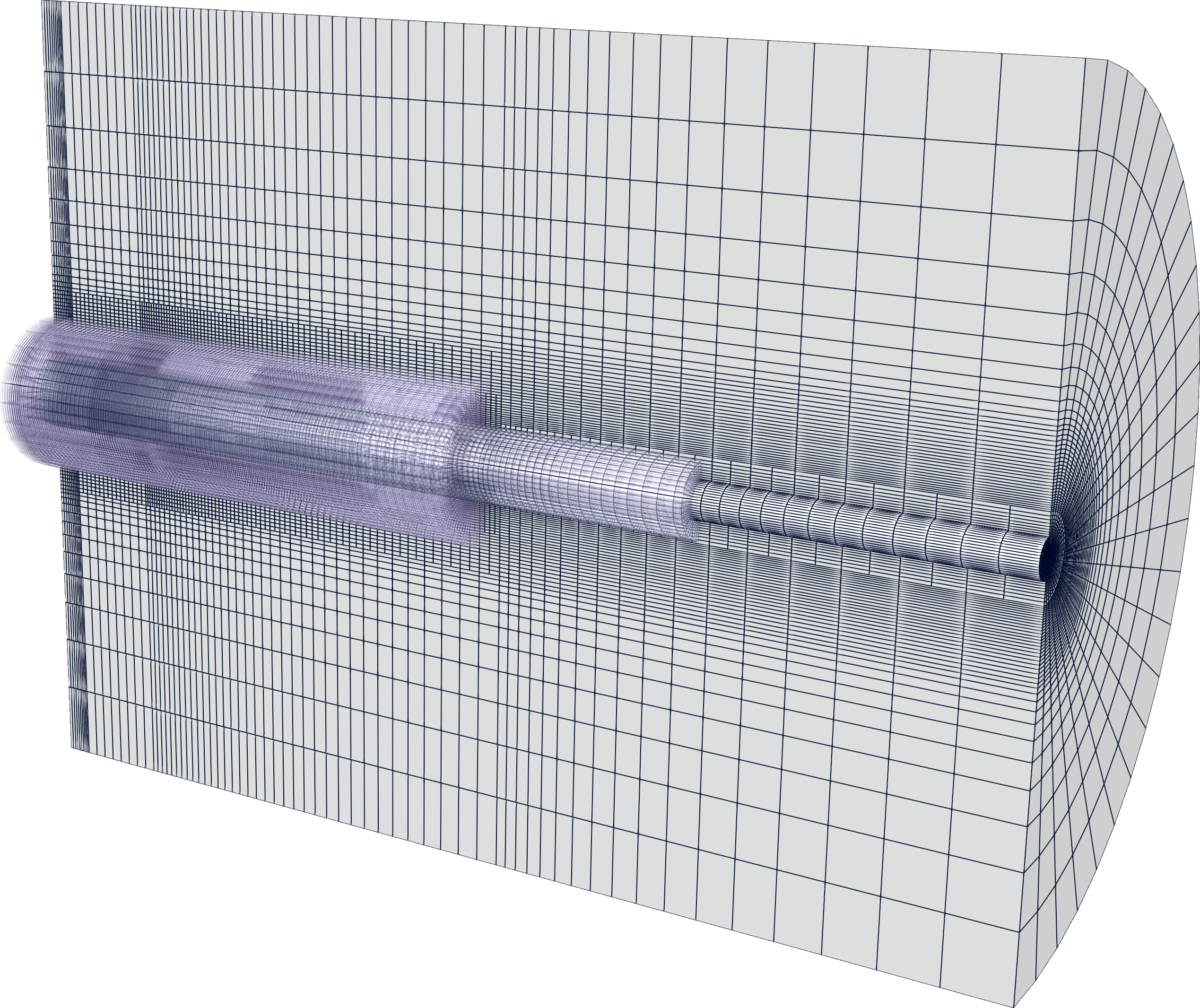}
		\caption{3D View}
		\label{subfig:mesh_visu_3D}
	\end{subfigure}
	\caption{Computational domain for the $\text{H}_2$ jet simulation. The dimensions of the setup are provided in (a) and
	a cut through the non-conforming 3D grid is visualized in (b).}
	\label{fig:mesh_visu}
\end{figure}
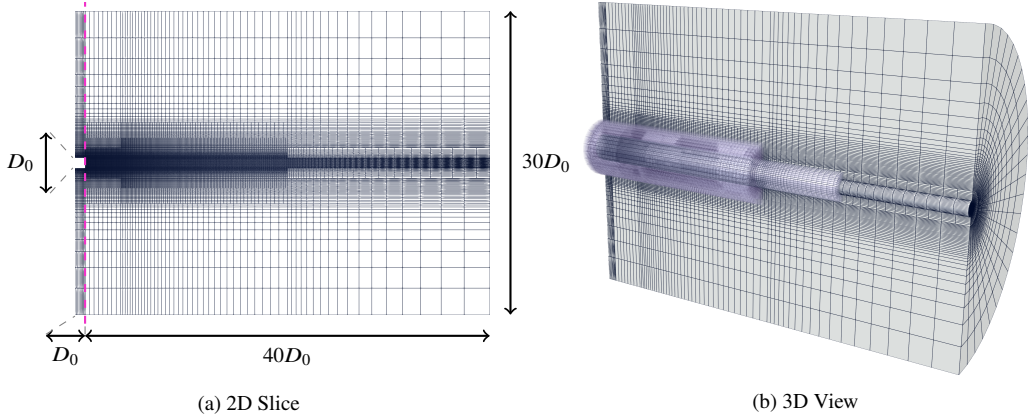

To demonstrate the ability of the presented hp-adaptive scheme to handle multi-component jet flows, we chose 
a benchmark similar to Hamzehloo et al.~\cite{Hamzehloo2014}, which examines an under-expanded hydrogen jet.
The setup employs a nozzle pressure ratio (NPR) of $10$ and is based on an experimental study of Ruggles and Ekoto~\cite{Ruggles2012}. 
The jet is simulated within a cylindrical domain with a diameter of $30D_0$ and a length of $41D_0$, where $D_0=\SI{1.5}{mm}$ denotes the nozzle diameter. 
While the study in~\cite{Hamzehloo2014} considers a high-pressure reservoir connected to the cylindrical domain via a convergent nozzle, 
we simplify the geometry by imposing a Dirichlet inlet boundary condition on a cross section of $D_0=\SI{1.5}{mm}$, using the state 
$\uB(\xB=\xB_{\text{inlet}},t=0)$ defined as 
\begin{equation}
    \begin{split}
		\uB(\xB=\xB_{\Omega},t=0)  &= \left(\rho,v_1,v_2,v_3,T,Y_1,Y_2 \right)^T \\
		&= \left(1.1579,0.0,0.0,0.0,296,1,0      \right)^T.
        \label{eq:air_jet_ics}
    \end{split}
\end{equation}
The inlet is extended by one nozzle diameter $D_0$ into the cylindrical domain. Boundaries surrounding the 
protruding inlet and the back wall of the cylindrical domain are treated as Euler slip walls. All remaining boundaries
are modeled with supersonic outflow conditions. Initially, a state  
\begin{equation}
    \begin{split}
        \uB(\xB=\xB_{\text{inlet}},t=0) &= \left(\rho,v_1,v_2,v_3,T,Y_1,Y_2 \right)^T \\ 
		&=\left(0.5115,1195.15,0.0,0.0,245.6,0,1\right)^T    
		\label{eq:h2_jet_ics}
    \end{split}
\end{equation}
is imposed within the domain $\Omega$. Table \ref{table:material_parameter} lists the material parameters 
employed for the $\text{H}_2$-jet simulation.
\renewcommand{\arraystretch}{1.5} 
\begin{table}[t]
	\centering
	\begin{tabular}{c|c|c|c|c|c|c}
		species&$\YB$&$\kappa$&$\nu \left[\frac{\SI{}{kg}}{\SI{}{m}\cdot\SI{}{s}}\right]$& $D \left[\frac{\SI{}{m}^2}{\SI{}{s}}\right]$ &Pr&$R \left[\frac{\SI{}{J}}{\SI{}{Kg}\cdot\SI{}{K}}\right]$ \\
		\hline
        \rule{0pt}{2ex}   
		Air          & $\left(1,0\right)^T$& 1.4  & 1.73E-5 & 1.E-6 & 0.66328  & 287.0  \\  
		$\text{H}_2$ & $\left(0,1\right)^T$& 1.41 & 0.84E-5 & 1.E-6 & 0.64301  & 4124.0 \\  
		\hline                                                   
	\end{tabular}
	\caption{Material parameters for air and $\text{H}_2$}
	\label{table:material_parameter}	
\end{table}
As depicted in Figure \ref{fig:mesh_visu}, the computational domain $\Omega$ is discretized with a block-structured, curved, non-conforming hexahedral 
mesh with roughly $700$ thousand elements. 
In smooth regions, the split-form DGSEM operator is employed with ansatz degrees between $N\in[2,4]$, while shocks and sharp concentration gradients
are stabilized with the \FV operator on a sub-cell grid of $N_{\text{FV}}=7$ sub-cells per direction. 
Positive density and pressure are ensured with the addition of the positivity preserving limiter of Zhang and Shu~\cite{Shu2010}. 
To avoid grid aligned shock instabilities at the curved Mach-disks, numerical fluxes between \FV elements and mixed \DGFV interfaces are computed with
a local Lax-Friedrichs solver. The Lax-Friedrichs solver was chosen over grid-aligned shock stabilization
techniques proposed by Fleischmann et al.~\cite{Fleischmann2020} and Chen et al.~\cite{Chen2020}, 
since these methods require an additional tuning parameter.

The setup is computed for $t\in(0,\SI{500}{\mu s}]$
on $N_\text{procs}=16384$ processors of the high performance computing (HPC) cluster \textit{HAWK}, requiring a total of $911$ thousand CPU hours. 
Load imbalances due to the variable ansatz degree and the sub-cell operator are dynamically balanced every $\mathcal{S}_{\text{DLB}}=50$ timesteps, whenever an imbalance
threshold of $\mathcal{T}_\text{DLB}=1.0$ is exceeded. For a detailed description and scaling analysis of the dynamic load balancing scheme, we refer to~\cite{Mossier2023}.
In total, $4357$ load balancing steps were performed during the simulation, amounting to $3.86\%$ percent of the total compute time. 

\subsection{Computational Results}

\begin{figure}[h!]
	\tikzsetnextfilename{results_t162}
	\input{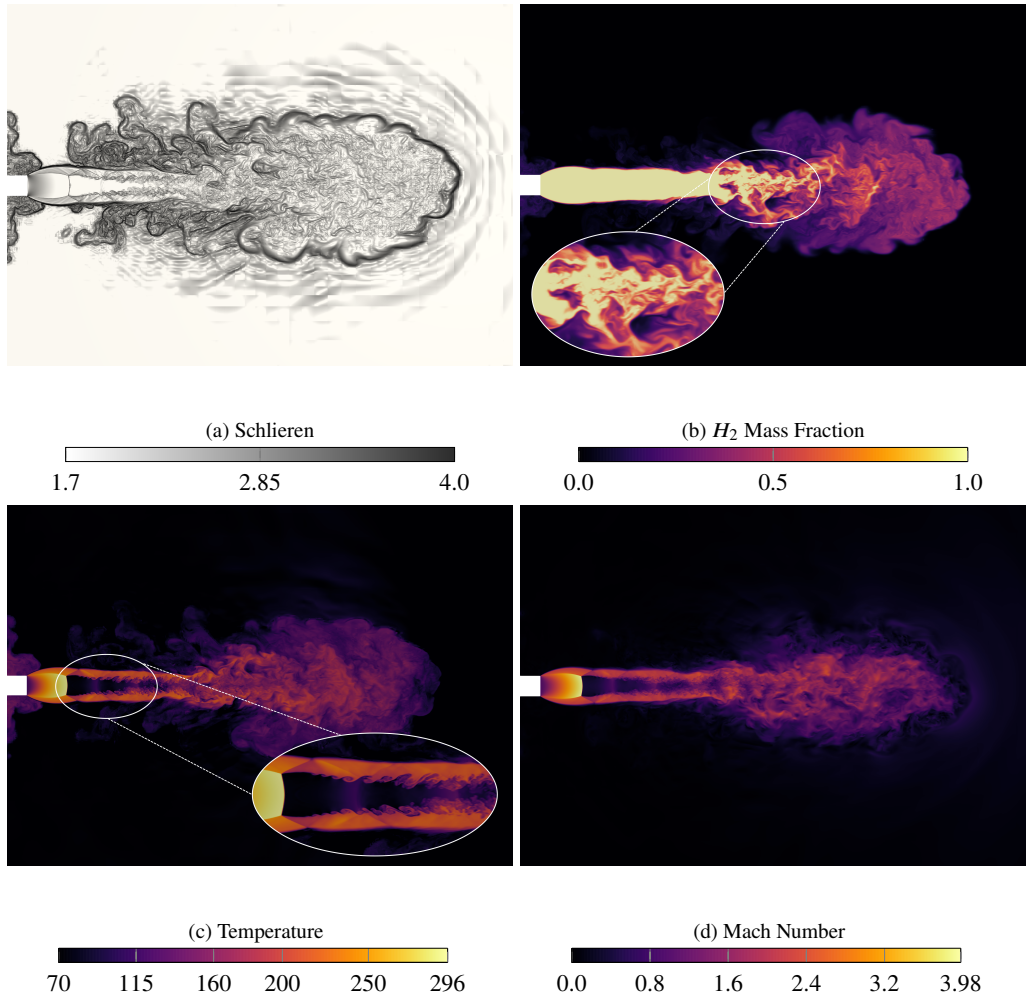}
	\caption{2D slice visualizing the flow field of the under-expanded $\text{H}_2$-jet in air at $t\approx161t^*$ with 
	a Schlieren plot (a), the mass fraction (b), the temperature field (c) and the Mach number (d).}
	\label{fig:h2_jet_results}
\end{figure}

The evaluation of the $\text{H}_2$-jet simulation focuses on three central aspects: the performance of the dynamic hp-refinement, 
an assessment of the effects of dynamic load balancing and a discussion of the flow field and jet-geometry in comparison to reference data. 
Following~\cite{Vuorinen2013}, we consider a non-dimensionalized timescale $t^*=\frac{D_0}{2 v_{\text{nozzle}}}$ for the 
under-expanded jet flow in terms of the nozzle diameter $D_0$ and the flow velocity at the nozzle exit, $v_\text{nozzle}$.
Figure \ref{fig:h2_jet_results} shows a slice of the domain at $t\approx161t^*$ and analyses the flow
field with a Schlieren image, a depiction of the $\text{H}_2$ mass fraction, the temperature $T$ and the Mach number. 

\begin{figure}[h!]
	\tikzsetnextfilename{3D_demo_t162}
	\input{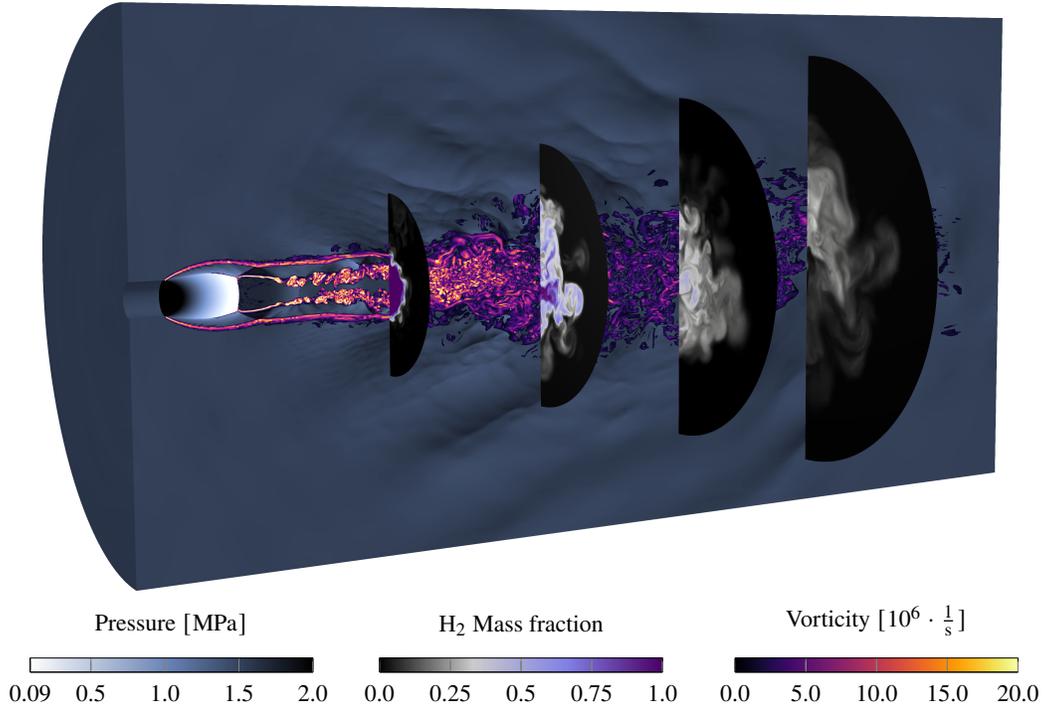}
	\caption{3D illustration of the jet flow at $t\approx805t^*$. 
	The plot shows the pressure distribution $p$, vorticity above a threshold $\omega>0.8\cdot10^6$ and 
	the $\text{H}_2$ mass fraction perpendicular to the main direction of propagation.}
	\label{fig:jet_3D}
\end{figure}

In addition, Figure \ref{fig:jet_3D} shows a cut through the domain at $t\approx805t^*$. 
The plot evaluates the pressure $p$ and superimposes the vorticity $\omega$ above a threshold of $\omega>0.8\cdot10^6$.
Furthermore, cross sections of the jet flow are depicted with slices perpendicular to jet propagation. 
Here, the $\text{H}_2$ mass fraction is evaluated to indicate the width of the jet.  

A comparison to results published by Hamzehloo et al.~\cite{Hamzehloo2014} shows a good qualitative agreement of the overall flow field.
This is supported by an analysis of the position and size of the first Mach-disk and the penetration of the jet at $t\approx161t^*$
in Table \ref{table:h2_jet_dims}, where a close qualitative match to the data of~\cite{Hamzehloo2014} is achieved. 

\begin{table}[b!]
	\centering
	\begin{tabular}{c|c|c|c|c}
		   & NPR &  \parbox{2.2cm}{\centering Mach-Disk  \\ Height [mm]} & 
		   			\parbox{2.2cm}{\centering Mach-Disk  \\  Width [mm]} & 
		   			\parbox{3.8cm}{\centering Centerline Penetration \\  $Z_\text{tip}$ [mm]} \\  
		\hline
        \rule{0pt}{2ex}   
		Present Study  & 10 & $3.10$	& $1.30$    & $31.40$ \\ 
		Hamzehloo      & 10 & $3.09$	& $1.34$ 	& $29.70$ \\ 
		\hline                                                   
	\end{tabular}
	\caption{Jet geometry comparison between the present study and that of Hamzehloo et al.~\cite{Hamzehloo2014} at $t\approx161t^*$}
	\label{table:h2_jet_dims}	
\end{table}

Further, the high-order approach employed in this study is able to capture significantly finer flow scales, especially within the jet
due to the low numerical dissipation of the DGSEM. 
This is particularly evident in the rapid development of eddies within the entropy shear layer downstream of the 
first Mach disk at $t\approx161t^*$. Consequently, the flow undergoes 
transition to a turbulent flow earlier within the jet than in the reference solution, promoting faster mixing of $\text{H}_2$ and air.

\begin{figure}[t]
	\centering
	\includegraphics[width=0.7\textwidth]{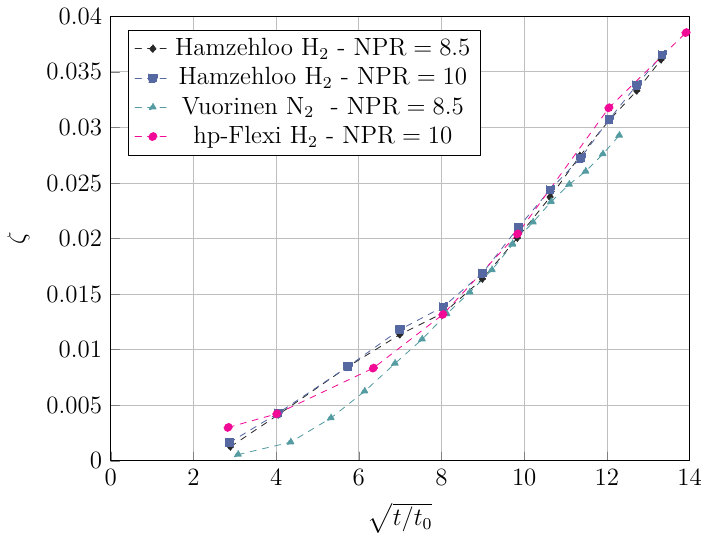}
	\caption{Normalized centerline jet-tip penetration $\mathcal{Z}_\text{tip}^*$ over the non-dimensional time $t^*$. The present study is compared
	against the results for $\text{H}_2$-jets of Hamzehloo et al.~\cite{Hamzehloo2014} and an $\text{N}_2$-jet of Vuorinen et al.~\cite{Vuorinen2013}.}
	\label{fig:centerline_depth}
\end{figure}

For a quantitative analysis of the jet-penetration, we compared the centerline penetration over time to data from 
Hamzehloo et al.~\cite{Hamzehloo2014} and a $\text{N}_2$-jet of Vuorinen et al.~\cite{Vuorinen2013} in Figure \ref{fig:centerline_depth}.
As discussed in~\cite{Vuorinen2013} a normalized jet-tip penetration $\smash{\mathcal{Z}_\text{tip}}^*$ can be obtained as 
\begin{equation}
	\mathcal{Z}_\text{tip}^*=\frac{Z_\text{tip}}{\sqrt[4]{\frac{\rho_0}{\rho_{\infty}}}}\propto \sqrt{\frac{t}{t^*}},
\end{equation}
with the nozzle density $\rho_0$ and the density in the undisturbed plenum $\rho_{\infty}$. Since the present setup
initiates the jet flow through an inflow condition, rather than an actual nozzle, the value for $\rho_0$ was extracted from the data of~\cite{Hamzehloo2014}.
$\smash{\mathcal{Z}_\text{tip}^*}$ is roughly proportional to $\smash{\sqrt{t/t^*}}$, leading to a near linear relation of $\smash{\mathcal{Z}_\text{tip}^*}$ over $\smash{\sqrt{t/t^*}}$.
For $\smash{\sqrt{t/t^*}\ge 8}$, the present study is in close agreement with the reference results. 

While the present computations predict a higher jet-tip penetration at the initial stages of the simulation $\sqrt{t/t^*}\approx 2.8$,
our data lies in between the curves of Vuorinen and Hamzehloo for $2.8<\sqrt{t/t^*}<8$.
The initial overestimation of $\mathcal{Z}_\text{tip}^*$ could be the consequence of modelling the inflow using boundary conditions instead of 
simulating an actual nozzle. The following phase of a slighly underpredicted jet-tip penetration could be explained by the faster development of turbulent
structures due to less numerical viscosity, leading to an earlier turbulent breakup. 

\begin{figure}[t]
	\tikzsetnextfilename{disc_dlb_t162}
	\input{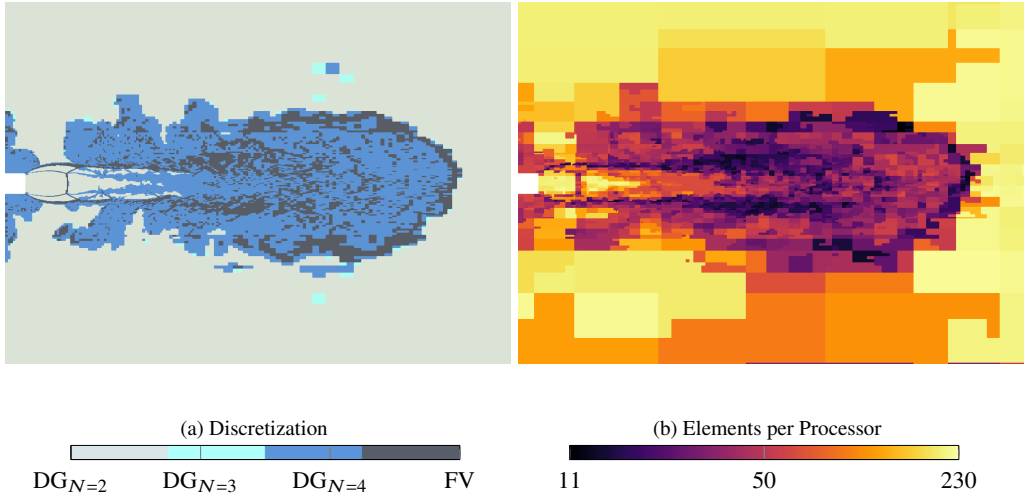}
	\caption{Depiction of the hp-adaptive element-local discretization and domain-decomposition in a slice of the $\text{H}_2$-jet simulation at $t^*=161$.
	In (a), a snapshot of the FV sub-cell element distribution and the local ansatz degree of the p-adaptive DG operator is shown. Figure (b) 
	illustrates the decomposition of the domain into sub-partitions for every processor. The color indicates the number of elements within 
	each partiton.}
	\label{fig:jet_hp_dlb}
\end{figure}

Finally, the hp-adaptive dynamic refinement and the dynamic load balancing are evaluated. Figure \ref{fig:jet_hp_dlb} 
provides a snapshot of the element-local discretization and the number of elements per processor at $t^*=161$. 
The indicators are demonstrated to perform as intended, with turbulent mixing zones resolved with the highest possible ansatz degree
and \FV sub-cells concentrated primarily at shocks and sharp concentration gradients. To mitigate the 
increased local computational costs caused by dynamic hp-refinement, DLB significantly reduces the number of elements per processor 
in the affected regions, as indicated in Figure \ref{subfig:jet_dlb}. 

\begin{figure}[htb!]
	\centering
	\includegraphics[width=1.0\textwidth]{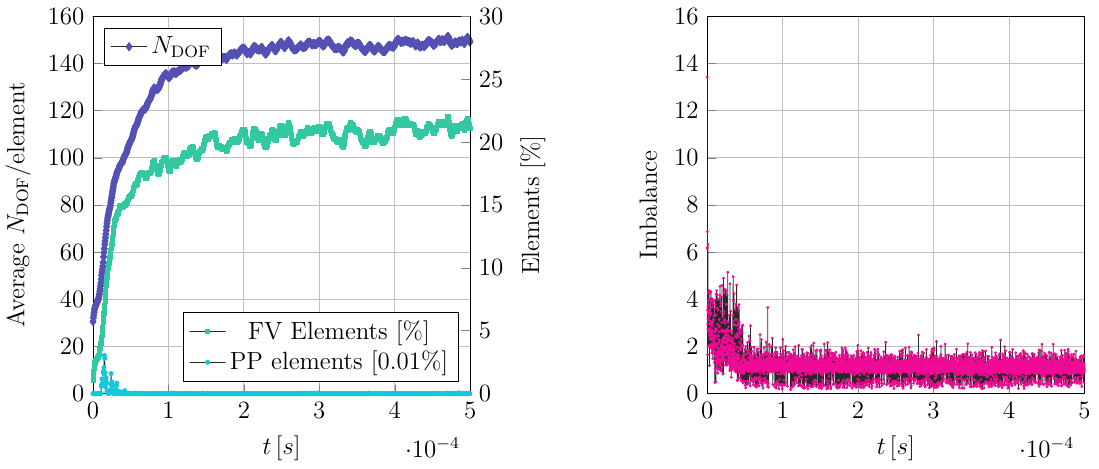}
	\caption{Statistics of the adaptive operator, PP-limiting and imbalance for the $\text{H}_2$-jet simulation.}
	\label{fig:jet_stats}
\end{figure}

Statistics on the element-local discretization over time and the number of elements affected by the PP-limiter are shown in figure  
\ref{fig:jet_stats}. Due to the nature of the non-conforming computational mesh, 
which allocates by design the majority of elements in the turbulent regions of the jet flow, 
the average number of DOFs and the proportion of \FV sub-cell elements are relatively high compared to the compressible TGV test 
case in Section \ref{subsec:tgv}. Still, a considerable degree of compression is maintained, 
as the resolution remains well below the maximum of $343$ DOFs per \FV sub-cell element.

As a final statistic, Figure \ref{fig:jet_stats} evaluates the imbalance over time $\mathcal{I}_\text{DLB}(t)$, where
$\mathcal{I}_\text{DLB}=0$ corresponds to an identical workload among all partitions as described in~\cite{Mossier2022}. 
Starting from an initial value of $\mathcal{I}_\text{DLB}>13$, the DLB scheme effectively reduces the imbalance to
roughly $\mathcal{I}_\text{DLB}\approx 1$. Although this still falls short of the optimal load distribution,
it constitutes a significant improvement in efficiency. 

\subsection{Parallel Performance Analysis}
\label{subsec:PrallelScaling}

\begin{figure}[b!]
	\centering
	\begin{subfigure}[t]{0.44\textwidth}	
		\vspace*{0pt}
		\includegraphics[width=1.0\textwidth]{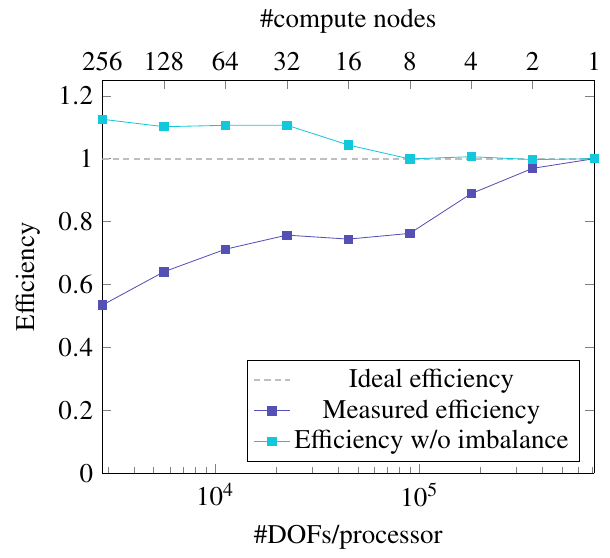}
		\caption{Parallel efficiency}
		\label{subfig:efficiency}
	\end{subfigure}
	\begin{subfigure}[t]{0.467\textwidth}	
		\vspace*{0pt}
		\includegraphics[width=1.0\textwidth]{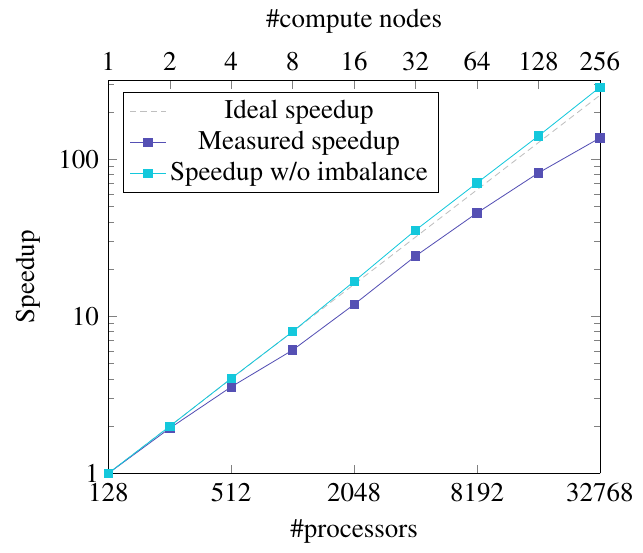}
		\caption{Parallel speedup}
		\label{subfig:speedup}
	\end{subfigure}
	\begin{subfigure}[t]{0.5\textwidth}	
		\vspace*{0pt}
		\includegraphics[width=1.0\textwidth]{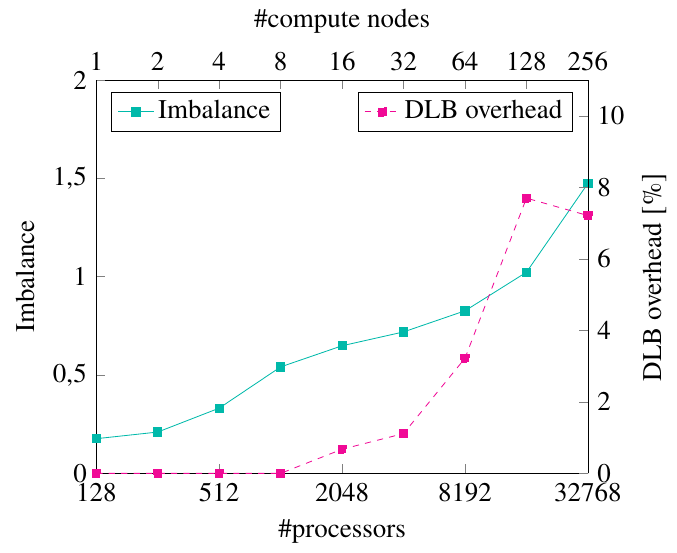}
		\caption{Imbalance $\mathcal{I}$ and DLB overhead}
		\label{subfig:Imbalance}
	\end{subfigure}
	\caption{Strong scaling results for the $\ce{H2}$-jet simulation over the time interval $t\in[100,101]\SI{}{\mu s}$ on up 
	to $256$ compute nodes of the HPC cluster \textit{LUMI}. The plots illustrate (a) parallel efficiency, 
	(b) parallel speedup, and (c) load imbalance $\mathcal{I}$ along with the computational overhead due of the load balancing procedure. 
	In (a) and (b), the theoretical efficiency and speedup, neglecting the measured imbalance, 
	are also shown to emphasize the impact of residual imbalance on scalability.}
	\label{fig:jet_scaling}
\end{figure}

For a quantitative assessment of the parallel performance, a strong scaling test is performed on the HPC cluster \textit{LUMI},
an HPE Cray EX system, featuring a high-speed HPE Slingshot $11$ interconnect and nodes equipped with 
$2\times\text{AMD}$ EPYC $64$-core processors and a memory of up to $1024$ GiB.
The scaling test recomputes the $\ce{H2}$-jet setup for the time interval $t\in[100,101]\SI{}{\mu s}$ 
on up to $256$ compute nodes, corresponding to $N_\text{procs}=32768$ processor units. The parallel efficiency and parallel speedup 
are evaluated in Figure \ref{subfig:efficiency} and Figure \ref{subfig:speedup} respectively, while the average imbalance 
and the DLB related overhead are analyzed in Figure \ref{subfig:Imbalance}. 

For up to $64$ nodes and a load of 
$\approx10^4$ DOFs per processor, the parallel efficiency surpasses $70\%$. 
When scaling to $256$ nodes, the per-processor load falls to $\approx3000$ DOFs, 
corresponding to an average of $22$ elements. While a significant degradation of parallel performance is observed here, the 
parallel efficiency still remaining above $50\%$. 
A comparison of the decline in parallel efficiency to the imbalance reveals a clear correlation between both trends.
Between $1$ and $8$  nodes, the imbalance rises significantly, followed by a more gradual increase between $8$ and $64$ nodes. Beyond $64$ nodes,
a sharp surge in the imbalance is apparent.

This behavior results from the chosen DLB scenario $\sigma$, where DLB is performed at most every
$\mathcal{S}_{\text{DLB}}=50$ time steps, provided that the imbalance $\mathcal{I}$ exceeds the threshold $\mathcal{T}_\text{DLB}=1.0$.
Due to this simple heuristic, DLB only takes effect during the scaling test for node counts above 8.

An analysis of the DLB overhead reveals that the selected heuristic is suboptimal. For up to $64$ nodes, the computation allocates 
less than $3\%$ of the wall clock time to DLB, while the parallel efficiency has already decreased by as much as $20\%$. This strongly suggests that 
increasing the frequency of DLB steps could yield significant benefits.

To motivate future improvements in the selection of a DLB heuristic and to better understand the causes of suboptimal
parallel scaling, we recalculated the theoretical parallel efficiency and speedup under ideal load balance by removing the imbalance. 
The theoretical results, without the influence of the imbalance, indicate ideal or even superlinear scaling similar to 
strong scaling results reported for the non-adaptive FLEXI implementation by Blind et al.~\cite{Blind2023}.

It is important to emphasize that the imbalance is measured and thus an approximation. Therefore, the theoretical scaling results
serve only as supporting evidence that the hp-adaptive operator itself exhibits good scaling, and that imbalance is the primary factor
responsible for the degradation of parallel efficiency.

\section{Conclusion}
\label{sec:Conclusion}
In this paper, we presented a novel hp-adaptive hybrid \DGFV discretization for compressible, turbulent multi-species flows. 
The scheme combines a p-adaptive DGSEM with an FV operator on an h-refined sub-grid to achieve high-order convergence in smooth regions
and accurate localization of shocks and material interfaces. It was implemented as an extension to the open-source code 
FLEXI and applied to the compressible Navier--Stokes equations with multiple species. 

Building on the adaptive hybrid \DGFV discretization strategy for hyperbolic gas- and droplet-dynamics of Mossier et al.~\cite{Mossier2022,Mossier2023,MossierPHD},
this work contributed key additions. First, the adaptive operator was extended to the hyperbolic-parabolic Navier--Stokes equations
by including viscous fluxes and a lifting procedure for second order terms. Further, the method was generalized to cover
non-conforming grids, greatly simplifying mesh generation. Since various combinations of operators and resolutions may be present between 
non-conforming element interfaces, the paper focuses on transforming the surface solutions and fluxes to conforming representations
to ensure consistent coupling. In addition, we address non-linear stability in the presence of under-resolved turbulence
by combining the adaptive scheme with the split-form DGSEM of Gassner et al.~\cite{Gassner2016}. Dynamic refinement is controlled 
by an error estimator, based on the decay rate of solution modes. In the present work, we promoted the idea of evaluating the indicator
on multiple variables to target specific flow features like shocks via the pressure, shear layers via the density and material interfaces via
the concentration. Since stability in multiple space dimensions can not be rigorously guaranteed for this hybrid DG/FV scheme,
the addition of a positivity preserving limiter was suggested.

The resulting scheme was validated with canonical free-stream preservation and experimental convergence tests. 
With an application to the compressible Taylor-Green vortex benchmark, we confirmed that the scheme is 
capable of distinguishing between shocks and under-resolved turbulence and achieves excellent results when compared
to other high-order methods in recent studies~\cite{Chapelier2024}.   
The novel indicator strategy, operating on multiple variables, was tested with a triplepoint shock interaction and demonstrated 
the desired behavior in detecting shear layers, material interfaces and shocks distinctly. 

Finally, the scheme was applied to a large-scale implicit large eddy simulation of a supersonic underexpanded hydrogen jet mixing with air. 
It demonstrated both computational efficiency and robustness on massively parallel high-performance systems, while
accurately capturing key physical features in close agreement with reference data from the literature. 

In the future, we integrate the hp-adaptive multi-species implementation with the sharp-interface 
framework for multi-phase flows with phase transition of J{\"o}ns et al.~\cite{Joens2023a} and Mossier et al.~\cite{Mossier2025}. 
Further, we will utilize the novel hp-adaptive hyperbolic-parabolic operator in sharp-interface simulations 
to explore compressible turbulence in multi-phase systems. 
Finally, in response to the shift of supercomputers toward graphics processing units, 
the hp-adaptive method will be integrated with the GPU-based implementation of FLEXI, published in~\cite{Kurz2025}.

\section*{Declarations}
\noindent \textbf{Funding} 
This work was funded by the European Union and has received funding from the European High Performance Computing Joint Undertaking (JU) 
and Sweden, Germany, Spain, Greece, and Denmark under grant agreement No 101093393.
Funding for this work was also received by the Deutsche Forschungsgemeinschaft (DFG, German Research Foundation) 
within the framework of the research unit FOR 2895 and FOR 2687.
Moreover, this research was funded by the DFG under Germany's Excellence 
Strategy EXC 2075-390740016 and the  GRK 2160/2, DROPIT. Further we want to gratefully acknowledge funded by the DFG through 
SPP 2410 Hyperbolic Balance Laws in Fluid Mechanics: Complexity, Scales, Randomness (CoScaRa). 
The simulations were performed on the national supercomputer 
HPE Apollo Systems \textit{HAWK} and the EuroHPC supercomputer \textit{LUMI}. 
We therefore gratefully acknowledge the support of the High Performance Computing Center Stuttgart (HLRS) for granting 
access to \textit{HAWK} under the grant number \textit{hpcmphas/44084} and 
the EuroHPC Joint Undertaking for awarding this project access to \textit{LUMI}
hosted by CSC (Finland) and the LUMI consortium through a EuroHPC Development Access call.
\\\\
\noindent \textbf{Conflict of interest} The corresponding author states on behalf of all authors, that there is no conflict of interest. 
\\\\
\noindent \textbf{Code availability} The open-source code FLEXI, on which all extensions are based, is available at www.flexi-project.org under the GNU GPL v3.0 license.
\\\\
\noindent \textbf{Availability of data and material} All data generated or analyzed during this study are included in this published article.

\bibliographystyle{spmpsci}     %
\bibliography{references.bib}   %

\end{document}